# High-frequency parametric approximation of the Floquet-Bloch spectrum for anti-tetrachiral materials

Andrea Bacigalupo[a], Marco Lepidi[b,*]

[a] IMT School for Advanced Studies Lucca, Piazza S. Francesco 19, 55100 Lucca (Italy)
[b] DICCA - Dipartimento di Ingegneria Civile, Chimica e Ambientale, Università di Genova, Via Montallegro 1, 16145 Genova (Italy)

## Abstract

The engineered class of anti-tetrachiral cellular materials is phenomenologically characterized by a strong auxeticity of the elastic macroscopic response. The auxetic behavior, accompanied by a marked anisotropy, is activated by rolling-up deformation mechanisms developed by the periodic pattern of stiff rings and flexible ligaments realizing the material micro-structure. In the absence of a soft matrix, a linear beam lattice model is formulated to describe the free dynamic response of the periodic cell. After a static condensation of the passive degrees-of-freedom, a general procedure is applied to impose the quasi-periodicity conditions of free wave propagation in the low-dimension space of the active degrees-of-freedom. The effects of different mechanical parameters on the band structure are analyzed by comparing the exact dispersion curves with explicit, although approximate, dispersion functions, obtained from asymptotic perturbation solutions of the eigenproblem rising up from the Floquet-Bloch theory. A general scheme for the formulation and solution of the perturbation equations is outlined, for the desired approximation order and depending on the dimension of the perturbation vector. A satisfying approximation accuracy is achieved for low-order asymptotic solutions in wide regions of the parameter space. The explicit dependence of the dispersion functions on the main parameters, including the cell aspect ratio, the ligament slenderness and the ring density, allows the in-depth discussion and – in a design perspective – the fine assessment of the spectrum properties, including specific features like wave velocities and band gap amplitudes.

## 1. Introduction

Auxetic materials possess the counter-intuitive and fascinating property to expand laterally when longitudinally stretched. Conversely, they contract laterally if compressed. This smart macroscopic behavior, mechanically described by negative Poisson's ratios, is seldom observable in nature, but artificially achievable by engineered materials [1–7]. Boosted by the rising demand for advanced applications in aerospace, chemical, naval, nuclear, biomedical, sport engineering, known artificial realizations of auxetic materials include polymeric or metallic foams and laminates [8, 9], as well as micro-structured composites, which typically possess periodic cellular geometries like – for instance – reticular networks, chiral lattices, re-entrant honeycombs and origami folds [10–16]. In respect to conventional materials, auxetic solids can offer complementary functional advantages, such as an increase of the shear modulus and fracture toughness, together with an increment of acoustic damping and indentation resistance [17–21]. Nowadays, one of the most promising theoretical and technological research challenges concern the employment of chiral auxetic media as versatile elastic guides for planar optical and acoustic waves. Indeed, in response to specific requirements, a proper tuning of the geometric, elastic and inertial properties, let these materials serve as efficient mechanical signal propagators (e.g. for information transfer), selective passive filters (e.g. for ambient noise reduction and vibration mitigation), low-cost integrity indicators (e.g. for damage assessment) [22–26].

In the class of chiral topologies, the anti-tetrachiral material is attracting major attention for its strong auxeticity, accompanied by a marked anisotropy of the elasto-dynamic response [12, 26–30]. The wave propagation properties of this material, including the effects of a soft cellular matrix embedding the micro-structure, have been studied according to the Floquet-Bloch theory [26]. The motivating key argument is that the geometric and mechanical periodic properties can be exploited as design parameters to control the band structure (Floquet-Bloch spectrum) in the frequency-wavevector $(\omega, \mathbf{k})$-space, in which the dispersion curves govern the harmonic content of the elastic waves traveling across the material [31]. Besides the traditional techniques of numerical continuation, which allow the direct treatment of the implicit spectrum function $F(\omega, \mathbf{k}) = 0$, an alternative approach may consist in seeking for an explicit – although approximate – parametric form $\omega = G(\mathbf{k})$ of each dispersion curve. In this respect, high-frequency asymptotic techniques have been applied in continuous models [32], perturbation methods have been formulated for multi-coupled mono-dimensional periodic systems [33, 34] and high-frequency homogenization have been proposed in micropolar generalized continua for chiral materials [35]. Among the others, a feasible objective of theoretical and applied interest relies in the rapid and efficient optimization of the material spectrum in terms of maximum amplification of the low-frequency band-gaps [36, 37].

*Corresponding author
*Email address:* marco.lepidi@unige.it (Marco Lepidi)

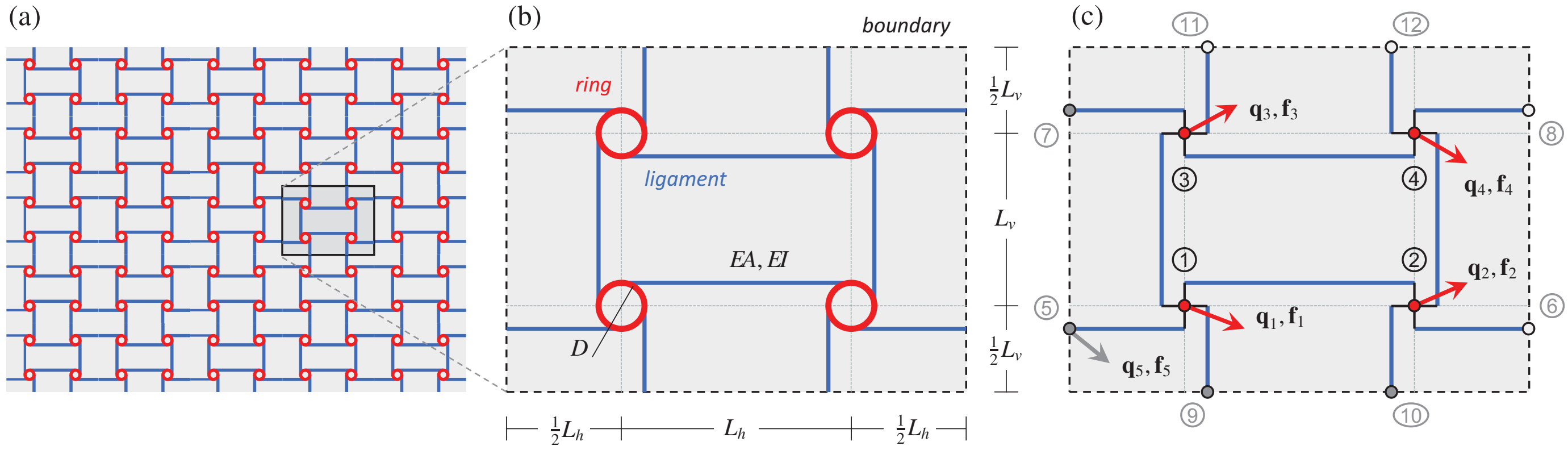

Figure 1: Anti-tetrachiral cellular material: (a) pattern, (b) periodic cell, (c) beam lattice model.

Within this framework, the present work focuses on the parametric analysis of the wave propagation properties (dispersion curves and polarization modes) of the beam lattice material with periodic anti-tetrachiral micro-structure. To this aim, a parametric linear model is formulated to describe the free undamped dynamic response of the periodic cell (Section 2). Then the eigenproblem governing the free propagation of planar harmonic waves is tackled (Section 3), by means of numerical solvers (exact solution) and different asymptotic perturbation methods (approximate solutions). The approximate solutions give explicit parametric functions which well-fit the optical and acoustic branches of the Floquet-Bloch spectrum, as well as the phase and group velocities (Section 3.3). A systematization effort is undertaken to present the perturbation equations and the related solution strategy in a formal mathematical scheme, suited for their generalization to different topologies of periodic beam lattice materials and their extension to the desired approximation order. Concluding remarks are finally pointed out.

## 2. Beam lattice model

A beam lattice model is formulated to describe the linear elasto-dynamic behavior of the rectangular elementary cell characterizing – at the microscopic scale of a two-dimensional domain – the cellular material featured by a periodic anti-tetrachiral geometry (Figure 1b). The internal structure, or *microstructure*, of the elementary cell is composed by four circular rings connected by twelve tangent ligaments. From an intuitive perspective, the auxetic behaviour can be physically justified by the opposite-sign, iso-amplitude rotations developed by any pair of adjacent disks in-a-row (or column), when the cell is stretched along one or the other ligament direction.

A rigid body model is adopted for the massive and highly-stiff rings, which are centered at the four corners of an ideal internal rectangle and possess identical mean diameter $D$. The free parameter $S$, denoting the ring width, allows the independent assignment of the rigid body mass $M$ and moment of inertia $J$. A linear, extensible, unshearable model of massless beam is employed for the light and flexible ligaments, in the small-deformation range. As long as the beam-ring connections realize perfectly-rigid joints, the natural length of the *inner* horizontal and vertical ligaments coincide with the width $L_h$ and height $L_v$ of the ideal internal rectangle, respectively. By virtue of the periodicity, the cell boundary crosses the midspan – and halves the natural length – of all the *outer* ligaments. Assuming the same linear elastic material (with Young's modulus $E$) and cross-section shape (with area $A$ and second area moment $I$) for each ligament, all the beams have identical extensional $EA$ and flexural rigidity $EI$. The effects of a homogeneous soft matrix, which may likely embed the microstructure [26], are neglected as first approximation.

Introducing a certain circular frequency $\Omega_r$ of the cellular solid as known dimensional reference, a suited minimal set $\mathbf{p}$ of independent nondimensional parameters, sufficient to describe the inertial, elastic and geometric properties of the model, is

$$\eta = \frac{L_h}{L_v}, \quad \delta = \frac{D}{L_h}, \quad \varrho^2 = \frac{I}{AL_h^2}, \quad \omega_c^2 = \frac{EA}{M\Omega_r^2 L_h}, \quad \chi^2 = \frac{J}{ML_h^2} \tag{1}$$

where the *aspect ratio* $\eta$ regulates the cell shape, by distinguishing the general rectangular shape ($\eta$ different from unity) from the particular square cell shape ($\eta = 1$). The $\delta$-parameter expresses the linear horizontal density of the circular rings, which can be interpreted as a measure of the material compositeness. The $\varrho$-parameter is the radius of gyration of the beam cross-section, divided by the beam length to assess the slenderness of the ligaments. Finally, $\chi^2$ accounts for the rotational-to-translational mass ratio of the disks, while $\omega_c$ is a nondimensional frequency suited to normalize the beam lattice spectrum.

### 2.1. Equations of motion

According to the mechanical assumptions and without additional mathematical approximations, the linear undamped free dynamics of the periodic cell is governed by a multi-degrees-of-freedom discrete model, referred to a set of twelve configuration nodes. The actual configuration of the i-th node (with $i = 1, ..., 12$) is described by three time-dependent components of non-rigid displacement, corresponding to the horizontal motion $U_i(t)$, the vertical motion $V_i(t)$ and the in-plane rotation $\phi_i(t)$, with respect to the position $\mathbf{x}_i = (X_i, Y_i)$ in the natural configuration. The nondimensional variables can be introduced

$$u_i = \frac{U_i}{L_h}, \qquad v_i = \frac{V_i}{L_h}, \qquad \tau = \Omega_r t \tag{2}$$

and $\mathbf{q} = (u_1, ..., u_{12}, v_1, ..., v_{12}, \phi_1, ..., \phi_{12})$ is the 36-by-one configuration vector collecting (column-wise) all the nondimensional displacements.

Depending on the position of the lumped masses in the discrete model and with reference to the labels in Figure 1c, the model nodes can conveniently be distinguished into

- four *internal* nodes located at the ring centroids (nodes ①...④), whose *active* displacements can be collected in the 12-by-one subvector $\mathbf{q}_a = (u_1, ..., u_4, v_1, ..., v_4, \phi_1, ..., \phi_4)$
- eight *external* nodes located at the midspan of the outer ligaments (nodes ⑤...⑫), whose *passive* displacements are collected in the complementary 24-by-one subvector $\mathbf{q}_p = (u_5, ..., u_{12}, v_5, ..., v_{12}, \phi_5, ..., \phi_{12})$

The distinction remarks that the internal nodes develop both nondimensional elastic $\boldsymbol{\sigma}_a$ *and* inertial forces $\mathbf{f}_a$, which actively participate in the dynamic cell equilibrium. On the contrary, the external nodes can develop only elastic forces $\boldsymbol{\sigma}_p$, which partially depend on the stiffness coupling with the internal nodes, and are requested to quasi-statically balance the reactive forces $\mathbf{f}_p$ transferred by the adjacent cells. Consistently, the active displacements $\mathbf{q}_a$ play the role of Lagrangian coordinates and suffice to describe the cell dynamics after a quasi-static condensation of the passive displacements $\mathbf{q}_p$.

According to displacement/force decomposition, the nondimensional equilibrium equation governing the undamped free oscillations of the discrete model has the matrix form

$$\begin{pmatrix} \mathbf{f}_a \\ \mathbf{0} \end{pmatrix} + \begin{pmatrix} \boldsymbol{\sigma}_a \\ \boldsymbol{\sigma}_p \end{pmatrix} = \begin{pmatrix} \mathbf{0} \\ \mathbf{f}_p \end{pmatrix} \tag{3}$$

or, making explicit the displacement and acceleration dependence of the elastic and inertial force,

$$\begin{bmatrix} \mathbf{M} & \mathbf{O} \\ \mathbf{O} & \mathbf{O} \end{bmatrix} \begin{pmatrix} \ddot{\mathbf{q}}_a \\ \ddot{\mathbf{q}}_p \end{pmatrix} + \begin{bmatrix} \mathbf{K}_{aa} & \mathbf{K}_{ap} \\ \mathbf{K}_{pa} & \mathbf{K}_{pp} \end{bmatrix} \begin{pmatrix} \mathbf{q}_a \\ \mathbf{q}_p \end{pmatrix} = \begin{pmatrix} \mathbf{0} \\ \mathbf{f}_p \end{pmatrix} \tag{4}$$

where dot indicates differentiation with respect to the nondimensional $\tau$-time and $\mathbf{O}$ stands for different-size empty matrices. According to a lumped mass description, non-null 12-by-12 mass submatrix $\mathbf{M}$ can be assumed diagonal. The symmetric 12-by-12 submatrix $\mathbf{K}_{aa}$ and 24-by-24 submatrix $\mathbf{K}_{pp}$ account for the stiffness of the internal and external nodes, respectively. The rectangular 12-by-24 submatrix $\mathbf{K}_{ap} = \mathbf{K}_{pa}^\top$ expresses the elastic coupling among the internal and external nodes (namely *global coupling*). If the active displacement vector is conveniently sorted and decomposed as $\mathbf{q}_a = (\mathbf{u}_a, \mathbf{v}_a, \boldsymbol{\phi}_a)$, with subvectors $\mathbf{u}_a = (u_1, ..., u_4)$, $\mathbf{v}_a = (v_1, ..., v_4)$ and $\boldsymbol{\phi}_a = (\phi_1, ..., \phi_4)$, the related submatrices $\mathbf{M}$ and $\mathbf{K}_{aa}$ read

$$\mathbf{M} = \frac{1}{\omega_c^2} \begin{pmatrix} \mathbf{I} & \mathbf{O} & \mathbf{O} \\ \mathbf{O} & \mathbf{I} & \mathbf{O} \\ \mathbf{O} & \mathbf{O} & \chi^2 \mathbf{I} \end{pmatrix}, \qquad \mathbf{K}_{aa} = \begin{pmatrix} \mathbf{K}_{uu} & \mathbf{O} & \mathbf{K}_{u\phi} \\ \mathbf{O} & \mathbf{K}_{vv} & \mathbf{K}_{v\phi} \\ \mathbf{K}_{\phi u} & \mathbf{K}_{\phi v} & \mathbf{K}_{\phi\phi} \end{pmatrix} \tag{5}$$

where the subsubmatrices $\mathbf{K}_{u\phi}$ and $\mathbf{K}_{v\phi}$ describe the rotational-translational coupling between the displacement components of the internal nodes (namely *local coupling*). The absence of any translational-translation coupling ($\mathbf{K}_{uv}$-subsubmatrix) can be noted. All the non-null entries of the stiffness matrices $\mathbf{K}_{aa}, \mathbf{K}_{pp}, \mathbf{K}_{ap}$ are reported in the AppendixA.

### 2.2. *Free wave propagation*

The passive displacement vector can be suitably decomposed $\mathbf{q}_p = (\mathbf{q}_p^-, \mathbf{q}_p^+)$, in order to separate the displacement components $\mathbf{q}_p^-$ of the left/bottom cell boundary $\Gamma^-$ (composed by the external nodes ⑤,⑦,⑨,⑩) from those of the right/top boundary $\Gamma^+$ (composed by the external nodes ⑥,⑧,⑪,⑫). A similar decomposition can be introduced for the vectors of the internal $\boldsymbol{\sigma}_p = (\boldsymbol{\sigma}_p^-, \boldsymbol{\sigma}_p^+)$ and external forces $\mathbf{f}_p = (\mathbf{f}_p^-, \mathbf{f}_p^+)$. Introducing a stiffness matrix partition consistent with this decomposition, the dynamic (upper) part of the equation (4) reads

$$\mathbf{M}\ddot{\mathbf{q}}_a + \mathbf{K}_{aa}\mathbf{q}_a + \mathbf{K}_{ap}^+\mathbf{q}_p^+ + \mathbf{K}_{ap}^-\mathbf{q}_p^- = \mathbf{0} \tag{6}$$

whereas the quasi-static (lower) part can be written

$$\begin{bmatrix} \mathbf{K}_{pa}^- \\ \mathbf{K}_{pa}^+ \end{bmatrix} \mathbf{q}_a + \begin{bmatrix} \mathbf{K}_{pp}^= & \mathbf{K}_{pp}^\mp \\ \mathbf{K}_{pp}^\pm & \mathbf{K}_{pp}^\# \end{bmatrix} \begin{pmatrix} \mathbf{q}_p^- \\ \mathbf{q}_p^+ \end{pmatrix} = \begin{pmatrix} \mathbf{f}_p^- \\ \mathbf{f}_p^+ \end{pmatrix} \tag{7}$$

and finally the $(j)$-th cell requires the boundary conditions of geometric compatibility and quasi-static equilibrium

$$\mathbf{q}_{p(j)}^- = \mathbf{q}_{p(j-1)}^+, \qquad \boldsymbol{\sigma}_{p(j)}^- = \boldsymbol{\sigma}_{p(j-1)}^+, \qquad \forall j \tag{8}$$

where the $(j-1)$-subscript refers to the leftward/downward adjacent cells, across the $\Gamma^-$ boundary.

The propagation of a one-dimensional free-vibration wave along a infinite periodic cellular domain requires the displacements in the generic $L$-long cell being equal to the corresponding displacements in the adjacent cells times $e^{\imath kL}$, where $k$ is the (dimensional) propagation constant, or *wavenumber*. Waves propagating in a two-dimensional domain are characterized by the *wavevector* $\mathbf{k} = (k_1, k_2)$. The displacement propagation is enabled by the cross-boundary exchange of non-null internal forces, obeying to the same exponential law.

According to the Floquet-Bloch theory, the following representations of the nodal displacements and forces are introduced

$$\mathbf{q}_a = \mathbf{B}_a\mathbf{p}_a, \qquad \mathbf{q}_p = \mathbf{B}_p\mathbf{p}_p, \qquad \boldsymbol{\sigma}_p = \mathbf{B}_p\boldsymbol{\tau}_p \tag{9}$$

where $\mathbf{p}_a$, $\mathbf{p}_p$, $\boldsymbol{\tau}_p$ are *auxiliary* vectors collecting the nodal displacements/forces in the $\mathbf{k}$-transformed space and the nondimensional transformation matrices read

$$\mathbf{B}_a = \begin{bmatrix} \mathbf{A}_4 & \mathbf{O} & \mathbf{O} \\ \mathbf{O} & \mathbf{A}_4 & \mathbf{O} \\ \mathbf{O} & \mathbf{O} & \mathbf{A}_4 \end{bmatrix}, \qquad \mathbf{B}_p = \begin{bmatrix} \mathbf{A}_8 & \mathbf{O} & \mathbf{O} \\ \mathbf{O} & \mathbf{A}_8 & \mathbf{O} \\ \mathbf{O} & \mathbf{O} & \mathbf{A}_8 \end{bmatrix} \tag{10}$$

where the submatrices

$$\mathbf{A}_4 = \mathrm{diag}\left(e^{\imath\mathbf{k}\cdot\mathbf{x}_1}, ..., e^{\imath\mathbf{k}\cdot\mathbf{x}_4}\right), \quad \mathbf{A}_8 = \mathrm{diag}\left(e^{\imath\mathbf{k}\cdot\mathbf{x}_5}, ..., e^{\imath\mathbf{k}\cdot\mathbf{x}_{12}}\right) \tag{11}$$

According to the boundary-based decomposition for the passive vectors, the transformation (9) can be expressed as

$$\mathbf{q}_p^- = \mathbf{B}_p^- \mathbf{p}_p^-, \quad \boldsymbol{\sigma}_p^- = \mathbf{B}_p^- \boldsymbol{\tau}_p^-, \quad \mathbf{q}_p^+ = \mathbf{B}_p^+ \mathbf{p}_p^+, \quad \boldsymbol{\sigma}_p^+ = \mathbf{B}_p^+ \boldsymbol{\tau}_p^+ \tag{12}$$

where the decomposed transformation matrices read

$$\mathbf{B}_p^- = \begin{bmatrix} \mathbf{A}_4^- & \mathbf{O} & \mathbf{O} \\ \mathbf{O} & \mathbf{A}_4^- & \mathbf{O} \\ \mathbf{O} & \mathbf{O} & \mathbf{A}_4^- \end{bmatrix}, \qquad \mathbf{B}_p^+ = \begin{bmatrix} \mathbf{A}_4^+ & \mathbf{O} & \mathbf{O} \\ \mathbf{O} & \mathbf{A}_4^+ & \mathbf{O} \\ \mathbf{O} & \mathbf{O} & \mathbf{A}_4^+ \end{bmatrix} \tag{13}$$

and the submatrices

$$\begin{aligned} \mathbf{A}_4^- &= \mathrm{diag}\left(e^{\imath \mathbf{k}\cdot\mathbf{x}_5}, e^{\imath \mathbf{k}\cdot\mathbf{x}_7}, e^{\imath \mathbf{k}\cdot\mathbf{x}_9}, e^{\imath \mathbf{k}\cdot\mathbf{x}_{10}}\right), \\ \mathbf{A}_4^+ &= \mathrm{diag}\left(e^{\imath \mathbf{k}\cdot\mathbf{x}_6}, e^{\imath \mathbf{k}\cdot\mathbf{x}_8}, e^{\imath \mathbf{k}\cdot\mathbf{x}_{11}}, e^{\imath \mathbf{k}\cdot\mathbf{x}_{12}}\right) \end{aligned} \tag{14}$$

Finally, the periodicity conditions governing the wave propagation, as they apply to the *auxiliary* vectors

$$\mathbf{p}_{p(j-1)}^+ = \mathbf{p}_{p(j)}^+, \qquad \boldsymbol{\tau}_{p(j-1)}^+ = \boldsymbol{\tau}_{p(j)}^+ \tag{15}$$

lead to the quasi-periodicity conditions applied to the (non-transformed) displacement/force vectors

$$\mathbf{q}_{p(j-1)}^+ = \mathbf{L}\mathbf{q}_{p(j)}^+, \qquad \boldsymbol{\sigma}_{p(j-1)}^+ = \mathbf{L}\boldsymbol{\sigma}_{p(j)}^+ \tag{16}$$

where $\mathbf{L}$ is the nondimensional *transfer* matrix

$$\mathbf{L} = \begin{bmatrix} \mathbf{D}_4 & \mathbf{O} & \mathbf{O} \\ \mathbf{O} & \mathbf{D}_4 & \mathbf{O} \\ \mathbf{O} & \mathbf{O} & \mathbf{D}_4 \end{bmatrix} \tag{17}$$

and the submatrices

$$\mathbf{D}_4 = \mathrm{diag}\left(e^{\imath \mathbf{k}\cdot\mathbf{d}_{56}}, e^{\imath \mathbf{k}\cdot\mathbf{d}_{78}}, e^{\imath \mathbf{k}\cdot\mathbf{d}_{911}}, e^{\imath \mathbf{k}\cdot\mathbf{d}_{1012}}\right) \tag{18}$$

where $\mathbf{d}_{ij} = \mathbf{x}_j - \mathbf{x}_i$ is an auxiliary vector pointing from the $i$-th to the $j$-th external nodes.

Recalling the boundary conditions (8) and replacing the internal with the external forces through the relations $\boldsymbol{\sigma}_{p(j)}^+ = \mathbf{f}_{p(j)}^+$ and $\boldsymbol{\sigma}_{p(j)}^- = -\mathbf{f}_{p(j)}^-$, a suited single-cell form of the quasi-periodicity condition (16) is achieved

$$\mathbf{q}_p^+ = \mathbf{L}\mathbf{q}_p^-, \qquad \mathbf{f}_p^+ = -\mathbf{L}\mathbf{f}_p^- \tag{19}$$

where the ubiquitous $j$-subscript has been omitted.

To the specific purposes of the present work, it is necessary to distinguish the left $\Gamma_l^-$ (external nodes ⑤,⑦) from the bottom cell boundary $\Gamma_b^-$ (nodes ⑨,⑩), as well as the right $\Gamma_r^+$ (nodes ⑥,⑧) from the top cell boundary $\Gamma_t^+$ (nodes ⑪,⑫). Introducing a further decomposition of the passive displacement vectors according to this distinction, the equations (19) reads

$$\begin{pmatrix} \mathbf{q}_r^+ \\ \mathbf{q}_t^+ \end{pmatrix} = \begin{bmatrix} e^{\imath\beta_1}\mathbf{I} & \mathbf{O} \\ \mathbf{O} & e^{\imath\beta_2}\mathbf{I} \end{bmatrix} \begin{pmatrix} \mathbf{q}_l^- \\ \mathbf{q}_b^- \end{pmatrix}, \tag{20}$$

$$\begin{pmatrix} \mathbf{f}_r^+ \\ \mathbf{f}_t^+ \end{pmatrix} = \begin{bmatrix} -e^{\imath\beta_1}\mathbf{I} & \mathbf{O} \\ \mathbf{O} & -e^{\imath\beta_2}\mathbf{I} \end{bmatrix} \begin{pmatrix} \mathbf{f}_l^- \\ \mathbf{f}_b^- \end{pmatrix} \tag{21}$$

where the nondimensional propagation constants, or *wavenumbers* $\beta_1 = \mathbf{k}\cdot\mathbf{d}_{56} = \mathbf{k}\cdot\mathbf{d}_{78}$ (along the horizontal direction) and $\beta_2 = \mathbf{k}\cdot\mathbf{d}_{911} = \mathbf{k}\cdot\mathbf{d}_{1012}$ (along the vertical direction) can be collected in the nondimensional wavevector $\mathbf{b} = (\beta_1, \beta_2)$. Moreover, it can be demonstrated that the wavevectors $\mathbf{b}_1 = (\beta_1, 0)$ and $\mathbf{b}_2 = (0, \beta_2)$, as defined in the periodic cell, identify the two orthogonal directions (spanned by the wavenumbers $\beta_1$ and $\beta_2$) traveled by the waves propagating along the orthotropy axes of the first-order equivalent continuum model of the material [26].

The free-wave propagation conditions (19) can be introduced in the quasi-static equation (7) to reduce the number of passive displacements to be condensed. Therefrom, solving the equation (7) in either the passive displacements $\mathbf{q}_p^-$ or the passive forces $\mathbf{f}_p^-$, both these variables are found to quasi-statically depend on the active displacements $\mathbf{q}_a$ through the relations

$$\mathbf{q}_p^- = \mathbf{R}(\mathbf{K}_{pa}^+ + \mathbf{L}\mathbf{K}_{pa}^-)\mathbf{q}_a, \tag{22}$$

$$\mathbf{f}_p^- = (\mathbf{K}_{pa}^- + (\mathbf{K}_{pp}^= + \mathbf{K}_{pp}^{\mp}\mathbf{L})\mathbf{R}(\mathbf{K}_{pa}^+ + \mathbf{L}\mathbf{K}_{pa}^-))\mathbf{q}_a \tag{23}$$

where it can be verified that the $\mathbf{b}$-dependent auxiliary matrix $\mathbf{R} = -(\mathbf{L}\mathbf{K}_{pp}^{\mp}\mathbf{L} + \mathbf{L}\mathbf{K}_{pp}^{=} + \mathbf{K}_{pp}^{\#}\mathbf{L} + \mathbf{K}_{pp}^{\pm})^{-1}$ is diagonal.

Forcing the quasi-periodic condition (19) and the quasistatic relation (23) into the equation (6), the cell dynamics is fully described in the active displacement space, and is governed by the equation of motion

$$\mathbf{M}\ddot{\mathbf{q}}_a + \mathbf{K}_b\mathbf{q}_a = \mathbf{0} \tag{24}$$

where it can be demonstrated that the $\mathbf{b}$-dependent stiffness matrix $\mathbf{K}_b = \mathbf{K}_{aa} + (\mathbf{K}_{ap}^- + \mathbf{K}_{ap}^+\mathbf{L})\mathbf{R}(\mathbf{K}_{pa}^+ + \mathbf{L}\mathbf{K}_{pa}^-)$ is Hermitian.

Denoting $\omega$ the unknown nondimensional frequency, imposing the oscillatory solution $\mathbf{q}_a = \boldsymbol{\psi}_a e^{\imath\omega\tau}$ and eliminating the dependence on time, a non-standard eigenproblem in the unknown eigenvalues $\lambda$ and eigenvectors $\boldsymbol{\psi}_a$ can be stated

$$(\mathbf{K}_b - \lambda\mathbf{M})\boldsymbol{\psi}_a = \mathbf{0} \qquad \text{where} \qquad \lambda = \frac{\omega^2}{\omega_c^2} \tag{25}$$

which, decomposing the mass matrix in the form $\mathbf{M} = \mathbf{Q}^\top\mathbf{Q}$ (the decomposition is unique as the matrix $\mathbf{M}$ is diagonal), can conveniently be reduced to the standard form

$$(\mathbf{H} - \lambda\mathbf{I})\boldsymbol{\varphi}_a = \mathbf{0} \tag{26}$$

where the auxiliary eigenvector $\boldsymbol{\varphi}_a = \mathbf{Q}\boldsymbol{\psi}_a$ and the equivalent stiffness matrix $\mathbf{H} = \mathbf{Q}^{-\top}\mathbf{K}_b\mathbf{Q}^{-1}$ inherits the Hermitian property.

The eigenproblem solution gives twelve real-valued eigenvalues $\lambda_i$ (or frequencies $\omega_i$). It is worth remarking that, owing to the Hermitian property, the $\mathbf{H}$-matrix is certainly non-defective, that is, possesses a complete eigenspace spanned by twelve proper eigenvectors. Therefore, each eigenvalue $\lambda_i$ has coincident algebraic and geometric multiplicity $m_i$ and corresponds to a complex-valued eigenvectors $\boldsymbol{\psi}_{ai}$, collecting the active eigencomponents only. The passive eigencomponents depend on the active eigencomponents through the quasi-static $\beta$-dependent relations $\boldsymbol{\psi}_{pi}^- = \mathbf{R}(\mathbf{K}_{pa}^+ + \mathbf{L}\mathbf{K}_{pa}^-)\boldsymbol{\psi}_{ai}$ and $\boldsymbol{\psi}_{pi}^+ = \mathbf{L}\boldsymbol{\psi}_{pi}^-$.

## 3. FLOQUET-BLOCH SPECTRUM

Introducing the extended parameter vector $\boldsymbol{\mu} = (\mathbf{p}, \mathbf{b})$, each parameter-dependent eigenvalue $\lambda(\boldsymbol{\mu})$ can be regarded as one of the zeroes of the characteristic function $F(\lambda, \boldsymbol{\mu}) = \det(\mathbf{H}(\boldsymbol{\mu}) - \lambda \mathbf{I})$. Fixing certain structural properties, corresponding to an admissible point $\mathbf{p}$ of the physical space $\mathcal{P}$, all the eigenvalues can be determined under variation of the nondimensional wavevector $\mathbf{b}$ in the square Brillouin domain $\mathcal{D} = [-\pi, \pi] \times [-\pi, \pi]$. According to the Floquet-Bloch theory [31, 38, 39], this investigation – which in general has to be carried out numerically – can be limited to the range $\mathcal{B} = \mathcal{B}_1 \cup \mathcal{B}_2$, joining the edges $\mathcal{B}_1 = \{\beta_1 \in [0, \pi], \beta_2 = 0\}$ and $\mathcal{B}_2 = \{\beta_1 = 0, \beta_2 \in [0, \pi]\}$ which bound the irreducible zones of the $\mathcal{D}$-domain. The eigenvalue (or frequency) loci $\lambda_\beta$ in the $(\beta_i, \lambda)$-plane $\Pi$ constitute the so-called dispersion curves of the Floquet-Bloch spectrum for the waves propagating along the horizontal ($i = 1$) or vertical direction ($i = 2$).

The *roots* of the dispersion curves can be conventionally located in the origin of the $\mathcal{B}$-range (in $\beta_1 = \beta_2 = 0$), where the quasi-periodicity degenerates into standard periodicity. Here the eigenpairs can be interpreted as the natural (possibly null) frequencies and real-valued vibration modes of the single elementary cell in the free stationary oscillation regime of the periodic system. The natural modes can either be participated by all the active displacements (*global modes*) or, less often, dominated by a homogeneous subset of active displacements (*local modes*). Local modes can be classified as *translational modes*, if contributed by the horizontal or vertical degrees-of-freedom ($\mathbf{u}_a$ or $\mathbf{v}_a$, respectively), or *rotational modes*, if contributed by the rotation degrees-of-freedom ($\boldsymbol{\phi}_a$). This classification can also be extended to the $\beta$-dependent eigenvectors, which can be interpreted as polarization modes of the propagating wave, characterized by a certain (not null) wavenumber.

Parametric analyses, here not reported for the sake of conciseness, show that a generic parameter set $\boldsymbol{\mu}$ typically corresponds to well-distinct eigenvalues, whereas particular regions of the (extended) parameter space $\mathcal{M} = \mathcal{P} \times \mathcal{B}$ may correspond to multiple eigenvalues or to a cluster of close eigenvalues [40]. Borrowing the well-established nomenclature from the classic modal analysis, such *internal resonance* or *nearly-resonance* regions may be either confined at the boundaries of the $\mathcal{B}$-range or centered around particular $\mathbf{b}$-values. From the mechanical viewpoint, resonant regions are worth particular attention, since the interaction between identical or near eigenvalues can activate significant phenomena, such as crossing or veering of the frequency loci, localization or hybridization of the corresponding modal shapes, opening or closing of frequency band-gaps.

Local sensitivity analyses of the eigensolution can be conveniently performed by means of *asymptotic perturbation methods*, since such mathematical tools can furnish the sought eigenvalues and eigenvectors as explicit, through approximate, parametric functions. Paying the due algorithmic attention, the asymptotic perturbation solutions maintain their effectiveness and uniform validity also in the resonant regions [41, 42], where traditional techniques of numerical continuation tend to fail in individually tracking closely-spaced eigenvalue loci.

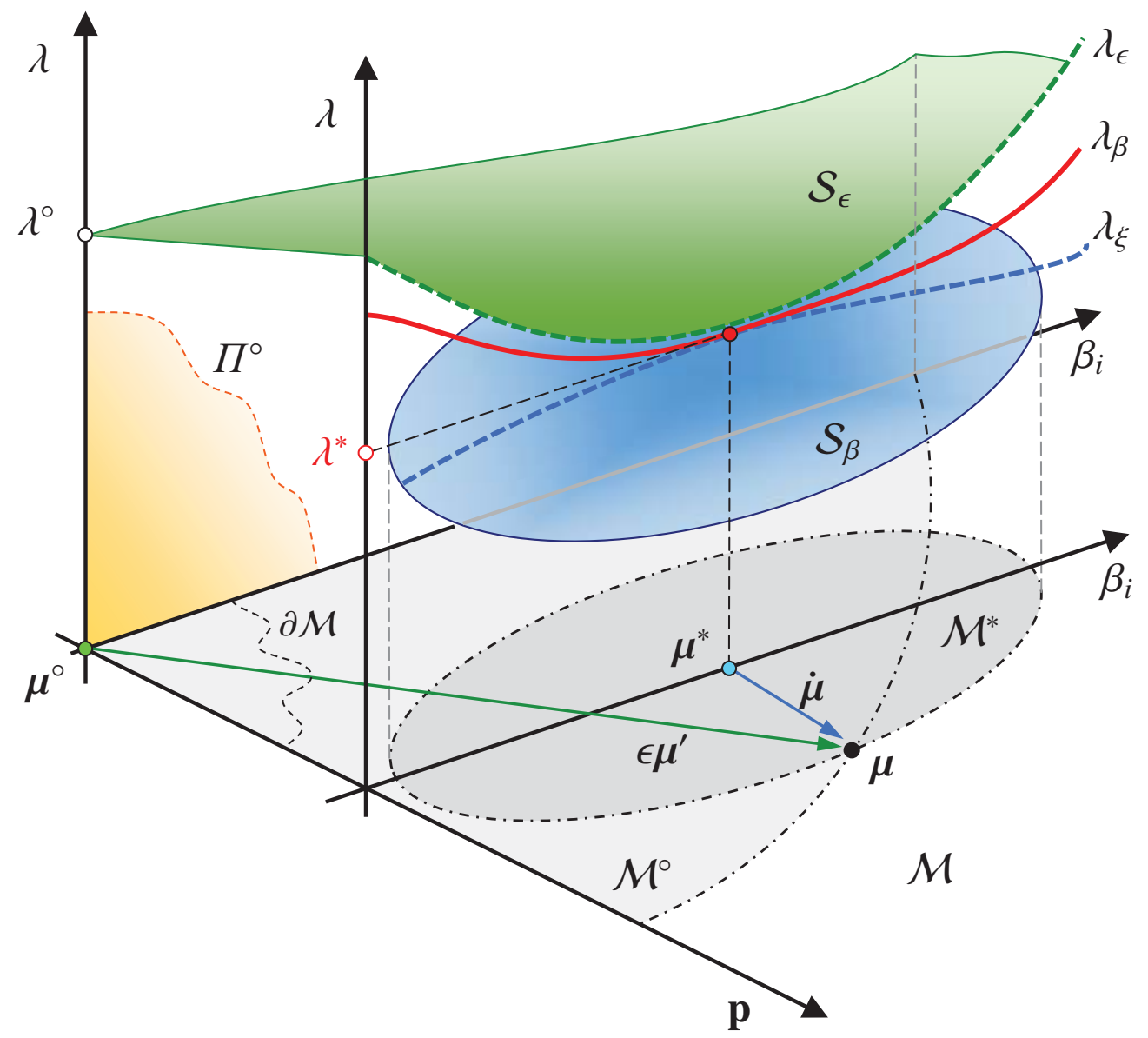

Figure 2: Qualitative schemes of the *SPPM* and *MPPM* for the asymptotic approximation of the exact eigenvalues

### 3.1. Single-parameter perturbation eigensolution

The asymptotic techniques are commonly based on selecting a certain point $\boldsymbol{\mu}^*$ of the parameter space $\mathcal{M}$. This preliminary choice fixes the wavenumbers and the structural properties of the reference cell, described by the matrix $\mathbf{H}^* = \mathbf{H}(\boldsymbol{\mu}^*)$. Even if not mandatory, a $\boldsymbol{\mu}^*$-selection which allows the explicit assessment of the (exact) eigenvalue $\lambda^*$ satisfying the equation $F(\lambda, \boldsymbol{\mu}^*) = 0$ is preferable. Taking the point $\boldsymbol{\mu}^*$ as reference, it is possible – in principle – to seek for an asymptotic eigenvalue approximation (the surface $\mathcal{S}_\beta$ in Figure 2) in the neighborhood $\mathcal{M}^*$ of $\boldsymbol{\mu}^*$, spanned by the perturbation vector $\dot{\boldsymbol{\mu}} = \boldsymbol{\mu} - \boldsymbol{\mu}^*$. More often, the asymptotic approximation is sought within the parameter subspace of major interest, that is, along one of the $\beta_i$-directions (with $i$=1,2) in the mono-dimensional neighborhood of the reference wavenumber $\beta_i^*$, spanned by the local abscissa $\xi = \beta_i - \beta_i^*$ [32]. Consequently, the only significant $\xi$-dependence is retained for the governing matrix $\mathbf{H}(\xi)$ and the characteristic function $F(\lambda, \xi)$. According to this simplification, in which $\xi$ acts as single perturbation parameter while the remaining parameter set $\mathbf{p}$ is freely assigned, this approach can be referred to as *Single-Parameter Perturbation Method* (*SPPM*).

Under the assumption of sufficient regularity, each exact eigenvalue of the matrix $\mathbf{H}(\xi)$ can tentatively be approximated by a series function $\lambda(\xi)$ of integer $\xi$-powers ($n \in \mathbb{N}$)

$$\lambda(\xi) = \lambda^* + \sum_n \lambda^{(n)} \xi^n = \lambda_0 + \dot{\lambda}\xi + \ddot{\lambda}\xi^2 + \ldots + \lambda^{(n)}\xi^n + \ldots \qquad (27)$$

where the $\mathbf{p}$-dependent coefficient $n!\lambda^{(n)}$ can be regarded as the *unknown* $n$-th $\xi$-derivative (for $\xi = 0$) of the exact but implicit eigenvalue function $F(\lambda, \xi) = 0$. Consistently, the curve $\lambda_\xi$ described by the approximate function $\lambda(\xi)$ is tangent to the exact dispersion curve $\lambda_\beta$ in $\beta_i = \beta_i^*$ (or $\xi = 0$), while the approximation accuracy is expected to decay for increasing $\xi$-values.

Once the series $\lambda(\xi)$ has been established, the characteristic function becomes a composite single-variable function $G(\xi) = F(\lambda(\xi), \xi)$, which admits the Taylor expansion in $\xi$-powers

$$G(\xi) = G^* + \sum_n \frac{G^{(n)}}{n!}\xi^n = G^* + \dot{G}\xi + \frac{\ddot{G}}{2}\xi^2 + ... + \frac{G^{(n)}}{n!}\xi^n + ... \quad (28)$$

where $G^* = F(\lambda^*, 0)$ is certainly null, as far as $\lambda^*$ is the exact, known eigenvalue for $\xi = 0$ by hypothesis. The generic higher-order coefficient $G^{(n)}$ can be recognized as the $n$-th $\xi$-derivative (for $\xi = 0$) of the function $G(\xi)$. It must be calculated through the recursive application of the chain rule for the differentiation of single-variable composite functions.

To specify, each coefficient $G^{(n)}$ is a complete $n$-degree polynomial of all the *unknowns* up to the $n$-th derivative $\lambda^{(n)}$. With reference to the $\xi$-power, the lowest-order coefficients read

$$\xi^1: \quad \dot{G} = \dot{\lambda}F^{(1,0)} + F^{(0,1)} \quad (29)$$

$$\xi^2: \quad \ddot{G} = 2\ddot{\lambda}F^{(1,0)} + (\dot{\lambda})^2 F^{(2,0)} + 2\dot{\lambda}F^{(1,1)} + F^{(0,2)} \quad (30)$$

where the synthetic notation $F^{(h,k)} = \left[\partial^h_\lambda \partial^k_\xi F(\lambda, \xi)\right]_{\xi=0}$ is adopted for the partial derivatives of the characteristic function.

As complementary original contribution, the Scott version of the Faà di Bruno's formula [43] – generalizing the chain rule to higher derivatives – has been manipulated to achieve a recursive form of the generic $n$-th coefficient

$$\xi^n: \quad G^{(n)} = \sum_{\mathcal{S}(h,k)} \frac{n!}{h+k} F^{(h,k)} \eta^{[n]}_{hk} \quad (31)$$

where the index set $\mathcal{S}(h,k) = (\, h, k \in [0, h+k = n])$, and the recursive term is a non-differential, polynomial function

$$\eta^{[n]}_{hk} = \frac{1}{n\,\lambda^*} \sum_{j=1}^{n} (j(h+k) + j - n)\, \lambda^{(j)}\, \eta^{[n-j]}_{hk} \quad (32)$$

has to be initialized with $\eta^{[0]}_{hk} = (\lambda^*)^h$.

The approximate characteristic function must be satisfied by zeroing each $\xi^n$-order coefficient $G^{(n)}$. Thus, a chain of $n$ ordered equations (*perturbation equations*) is obtained. Starting with the zeroth-order solution, given by the twelve known eigenvalues $\lambda_0$ (*generating solution*), each perturbation equation of the chain involves a single unknown, that is, one of the higher-order coefficients. Depending on the algebraic multiplicity $m^*$ of each $\lambda^*$-eigenvalue, two fundamental cases occur

- *Simple eigenvalue*: if $\lambda^*$ is a simple root ($m^* = 1$) for the equation $F(\lambda^*, 0) = 0$, then the coefficient $F^{(1,0)} \neq 0$. Hence, the $\xi^1$-order equation (29) is linear in the unknown $\dot{\lambda}$, the $\xi^2$-order equation (30) is linear in the unknown $\ddot{\lambda}$ and so on. Therefore, the cascade solution (null if the numerator vanishes) for the lowest order equations is

$$\xi^1: \quad \dot{\lambda} = -\frac{F^{(0,1)}}{F^{(1,0)}} \quad (33)$$

$$\xi^2: \quad \ddot{\lambda} = -\frac{(\dot{\lambda})^2 F^{(2,0)} + 2\dot{\lambda}F^{(1,1)} + F^{(0,2)}}{2F^{(1,0)}} \quad (34)$$

  and, by extension, the $\xi^n$-order equation allows the determination of the $n$-th coefficient $\lambda^{(n)}$.

- *Double (semi-simple) eigenvalue*: if $\lambda^*$ is a double root ($m^* = 2$) for the equation $F(\lambda^*, 0) = 0$, then the coefficient $F^{(1,0)} = 0$, but $F^{(2,0)} \neq 0$. Since $\lambda^*$ must be non-defective (semi-simple), it can be proved that $F^{(1,0)} = 0$ [44]. Consequently, the $\xi^1$-order equation (29) is trivially satisfied, but leaves $\dot{\lambda}$ undetermined. Such an indetermination is cleared by the $\xi^2$-order equation (30), which is a quadratic in the $\dot{\lambda}$-unknown only, since the null multiplier $F^{(1,0)}$ affects the other unknown $\ddot{\lambda}$. Thus, the lowest order equations give

$$\xi^1: \quad \dot{\lambda} \quad \text{is undetermined} \quad (35)$$

$$\xi^2: \quad \dot{\lambda}_\pm = -\frac{F^{(1,1)} \pm \sqrt{(F^{(1,1)})^2 - F^{(2,0)}F^{(0,2)}}}{F^{(2,0)}} \quad (36)$$

  where the eigensensitivity pair $\dot{\lambda}_\pm$ splits the double root $\lambda^*$ in two distinct eigenvalues $\lambda^* + \xi\dot{\lambda}_\pm + O(\xi^2)$. The splitting is postponed to higher-orders only if a vanishing discriminant occurs, in consequence of the particular subcase $(F^{(1,1)})^2 = F^{(2,0)}F^{(0,2)}$. In the general case, instead, the higher unknowns $\lambda^{(n)}_\pm$ are determined by linear $\xi^{n+1}$-order equations, solved for one or the other of the $\dot{\lambda}_\pm$-values.

When the *SPPM* is applied to the eigenproblem (26) for $\beta^*_i = 0$ (one of the left bounds of the $\mathcal{B}$-range), the generating solution includes cases of single and double $\lambda^*$-eigenvalues. Tables 1 and 2 report the solution scheme required to achieve a fourth-order approximation $\lambda_i(\xi) = \lambda^*_i + \dot{\lambda}_i\xi + \ddot{\lambda}_i\xi^2 + \dddot{\lambda}_i\xi^3 + \ddddot{\lambda}_i\xi^4 + O(\xi^5)$ of all the twelve eigenvalues ($i = 1, ..., 12$, one for each row). It is worth remarking that the scheme of rectangular cell ($\eta \neq 1$ in Table 1) differs from that of square cell ($\eta = 1$ in Table 2), due to a dissimilar set of single and double $\lambda^*$-eigenvalues (first and second columns), requiring the solution of the perturbation equations up to the fourth and sixth-order, respectively.

Table 1: Solution scheme of the *SPPM*-based perturbation equations for the rectangular cell ($\beta^*_i = 0$, $\eta \neq 1$) up to the forth order.

| $m^*$ | $\lambda^*$ | $\xi^1$ | $\xi^2$ | $\xi^3$ | $\xi^4$ | $\xi^5$ | $\xi^6$ |
|---|---|---|---|---|---|---|---|
| 2 | $\lambda^*_{1,2}$ | - | $\dot{\lambda}_{1,2}$ | - | $\ddot{\lambda}_1$<br>$\ddot{\lambda}_2$ | $\dddot{\lambda}_1$<br>$\dddot{\lambda}_2$ | $\ddddot{\lambda}_1$<br>$\ddddot{\lambda}_2$ |
| 1 | $\lambda^*_3$ | $\dot{\lambda}_3$ | $\ddot{\lambda}_3$ | $\dddot{\lambda}_3$ | $\ddddot{\lambda}_3$ | ... | ... |
| 1 | $\lambda^*_4$ | $\dot{\lambda}_4$ | $\ddot{\lambda}_4$ | $\dddot{\lambda}_4$ | $\ddddot{\lambda}_4$ | ... | ... |
| 1 | $\lambda^*_5$ | $\dot{\lambda}_5$ | $\ddot{\lambda}_5$ | $\dddot{\lambda}_5$ | $\ddddot{\lambda}_5$ | ... | ... |
| 1 | $\lambda^*_6$ | $\dot{\lambda}_6$ | $\ddot{\lambda}_6$ | $\dddot{\lambda}_6$ | $\ddddot{\lambda}_6$ | ... | ... |
| 1 | $\lambda^*_7$ | $\dot{\lambda}_7$ | $\ddot{\lambda}_7$ | $\dddot{\lambda}_7$ | $\ddddot{\lambda}_7$ | ... | ... |
| 1 | $\lambda^*_8$ | $\dot{\lambda}_8$ | $\ddot{\lambda}_8$ | $\dddot{\lambda}_8$ | $\ddddot{\lambda}_8$ | ... | ... |
| 1 | $\lambda^*_9$ | $\dot{\lambda}_9$ | $\ddot{\lambda}_9$ | $\dddot{\lambda}_9$ | $\ddddot{\lambda}_9$ | ... | ... |
| 1 | $\lambda^*_{10}$ | $\dot{\lambda}_{10}$ | $\ddot{\lambda}_{10}$ | $\dddot{\lambda}_{10}$ | $\ddddot{\lambda}_{10}$ | ... | ... |
| 1 | $\lambda^*_{11}$ | $\dot{\lambda}_{11}$ | $\ddot{\lambda}_{11}$ | $\dddot{\lambda}_{11}$ | $\ddddot{\lambda}_{11}$ | ... | ... |
| 1 | $\lambda^*_{12}$ | $\dot{\lambda}_{12}$ | $\ddot{\lambda}_{12}$ | $\dddot{\lambda}_{12}$ | $\ddddot{\lambda}_{12}$ | ... | ... |

Legend: "-" = undetermined, "..." = higher-order unknowns

Table 2: Solution scheme of the *SPPM*-based perturbation equations for the rectangular cell ($\beta_i^* = 0$, $\eta = 1$) up to the forth order.

| $m^*$ | $\lambda^*$ | $\xi^1$ | $\xi^2$ | $\xi^3$ | $\xi^4$ | $\xi^5$ | $\xi^6$ |
|---|---|---|---|---|---|---|---|
| 2 | $\lambda^*_{1,2}$ | - | $\dot\lambda_{1,2}$ | - | $\ddot\lambda_1$<br>$\ddot\lambda_2$ | $\dddot\lambda_1$<br>$\dddot\lambda_2$ | $\ddddot\lambda_1$<br>$\ddddot\lambda_2$ |
| 2 | $\lambda^*_{3,4}$ | - | $\dot\lambda_{3,4}$ | - | $\ddot\lambda_3$<br>$\ddot\lambda_4$ | $\dddot\lambda_3$<br>$\dddot\lambda_4$ | $\ddddot\lambda_3$<br>$\ddddot\lambda_4$ |
| 2 | $\lambda^*_{5,6}$ | - | $\dot\lambda_{5,6}$ | - | $\ddot\lambda_5$<br>$\ddot\lambda_6$ | $\dddot\lambda_5$<br>$\dddot\lambda_6$ | $\ddddot\lambda_5$<br>$\ddddot\lambda_6$ |
| 2 | $\lambda^*_{7,8}$ | - | $\dot\lambda_{7,8}$ | - | $\ddot\lambda_7$<br>$\ddot\lambda_8$ | $\dddot\lambda_7$<br>$\dddot\lambda_8$ | $\ddddot\lambda_8$<br>$\ddddot\lambda_8$ |
| 2 | $\lambda^*_{9,10}$ | - | $\dot\lambda_{9,10}$ | - | $\ddot\lambda_9$<br>$\ddot\lambda_{10}$ | $\dddot\lambda_9$<br>$\dddot\lambda_{10}$ | $\ddddot\lambda_9$<br>$\ddddot\lambda_{10}$ |
| 1 | $\lambda^*_{11}$ | $\dot\lambda_{11}$ | $\ddot\lambda_{11}$ | $\dddot\lambda_{11}$ | $\ddddot\lambda_{11}$ | ... | ... |
| 1 | $\lambda^*_{12}$ | $\dot\lambda_{12}$ | $\ddot\lambda_{12}$ | $\dddot\lambda_{12}$ | $\ddddot\lambda_{12}$ | ... | ... |

Legend: "-" = undetermined, "..." = higher-order unknowns

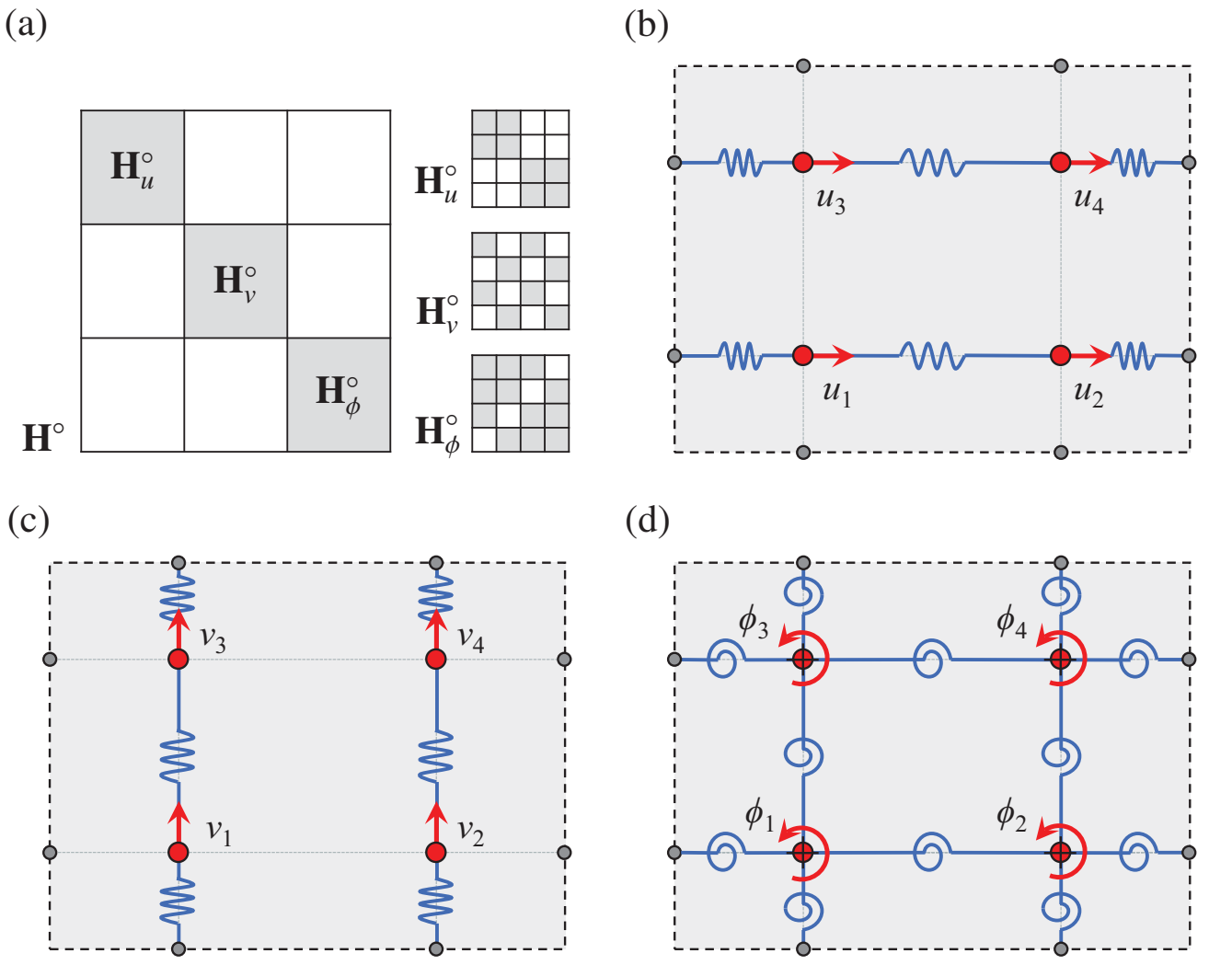

Figure 3: Physical realization of the ideal cell governed by the $\mathbf{H}^\circ$-matrix: (a) three-block diagonal $\mathbf{H}^\circ$-form, (b) $\mathbf{H}^\circ_u$-block subsystem, (c) $\mathbf{H}^\circ_v$-block subsystem, (d) $\mathbf{H}^\circ_\phi$-block subsystem.

### 3.2. *Multi-parameter perturbation eigensolution*

According to an alternative asymptotic technique, the eigenproblem (26) can be tackled by assigning a *suited* ordering to the full parameter set $\boldsymbol{\mu}$, in order to simultaneously assess the relative smallness of all its low-valued entries. Introducing a nondimensional auxiliary small parameter $\epsilon \ll 1$, the eigensolution around a particular $\beta_i^\circ$-point of the $\mathcal{B}$-range can be studied adopting the parameter ordering

$$\omega_c = \omega_c^\circ, \qquad \eta = \eta^\circ, \qquad \beta_i = \beta_i^\circ + \epsilon\,\beta_i' \qquad (37)$$
$$\delta = \epsilon\,\delta', \qquad \varrho = \epsilon\,\varrho', \qquad \chi = \epsilon\,\chi'$$

which is formally equivalent to order the parameter set $\boldsymbol{\mu}(\epsilon) = \boldsymbol{\mu}^\circ + \epsilon\boldsymbol{\mu}'$, where the *dominant* $O(1)$-part $\boldsymbol{\mu}^\circ$ is fully defined by the $(\omega_c^\circ, \eta^\circ, \beta_i^\circ)$-components, while the *perturbating* $O(\epsilon)$-part $\boldsymbol{\mu}'$ is attributed to the $(\beta_i', \delta', \varrho', \chi')$-components. Consistently, the matrix $\mathbf{H}^\circ = \lim_{\epsilon\to 0} \mathbf{H}(\boldsymbol{\mu}(\epsilon))$ corresponding to the $\boldsymbol{\mu}^\circ$-point is expected to capture the key-features of the wave dynamics.

As major formal difference with respect to the *SPPM*, the $\epsilon$-parameter plays the role of single perturbation parameter, but suffices to regulate the (small) amplitude of the multi-parametric perturbation $\boldsymbol{\mu}'$. Consequently, this approach can be conventionally referred to as *Multi-Parameter Perturbation Method* (*MPPM*). As substantial point, once a certain parameter ordering is chosen, the reference point $\boldsymbol{\mu}^\circ$ cannot be chosen arbitrarily in the $\mathcal{M}$-space, as $\boldsymbol{\mu}^\circ$ must be the asymptotic limit of $\boldsymbol{\mu}(\epsilon)$ for vanishing $\epsilon$. Consequently, the parameter combination $\boldsymbol{\mu}^\circ$ must be fixed in a restriction $\mathcal{M}^\circ$ of the $\mathcal{M}$-space, where only the non null $O(1)$-parameters can be freely assigned (Figure 2).

From the technical viewpoint, since $\boldsymbol{\mu}^\circ$ plays a merely algorithmic role as starting point of the perturbation analysis, it can sit on – or even lie outside – the boundary $\partial\mathcal{M}$ of the physically admissible (meaningful) region in the $\mathcal{M}$-space. Accordingly, the structural realization governed by the $\mathbf{H}^\circ$-matrix must be intended as a mathematical abstraction in the general case (including physically meaningless $\boldsymbol{\mu}^\circ$points). Therefore it will be referred to as *ideal cell* in the following. For the anti-tetrachiral material, a coherent mechanical interpretation of the ideal cell is possible, since the $\mathbf{H}^\circ$-matrix exhibits a three-block diagonal form (Figure 3a). As far as the diagonal blocks $\mathbf{H}^\circ_u$, $\mathbf{H}^\circ_v$, $\mathbf{H}^\circ_\phi$ are uncoupled from each other, they ideally govern three independent subsystems. Each subsystem can be demonstrated to possess a homogeneous (horizontal, vertical or rotational) set of four active degrees-of-freedom, with translational or rotational masses connected to each other by a stiffness coupling. This elastic coupling is described by the non-null entries of each block, and can be considered equivalent to internal linear springs joining pairs of adjacent nodes (Figure 3b,c,d). The ideal cell corresponds to the superposition, without interaction, of all the independent subsystems. From a rigorous perspective, it may be worth remarking that the uncoupled ideal subsystems are not obtained by trivially disassembling the original cell model into its different components, as could be done by merely neglecting the small coupling terms. On the contrary, they physically realize the asymptotic limit of the fully-coupled cell model for vanishing $\epsilon$-values, consistently with the perturbation scheme (see also [42, 45]). As minor remark, the ideal cell does not possess properties of anti-chirality, which rise up with the first order matrix perturbation.

As far as the Hermitian property holds for the $\mathbf{H}^\circ$-matrix, the eigenspectrum of the ideal cell still admits real-valued non-defective eigenvalues $\lambda^\circ$, satisfying the characteristic equation $F(\lambda^\circ, \boldsymbol{\mu}^\circ) = \det(\mathbf{H}^\circ - \lambda^\circ \mathbf{I})$. The eigenvalues of the ideal cell are discriminant for the suitability of the assigned parameter ordering [41]. Roughly, a suited ordering should simplify as much as possible the $\mathbf{H}^\circ$-spectrum by maximizing the algebraic multiplicity $m^\circ$ of each eigenvalue $\lambda^\circ$. Owing to the likely occurrence of multiple eigenvalues, the ideal cell can be regarded as a *perfectly tuned*, or simply *perfect* system, according to the well-established nomenclature of periodic structures. As operational criterion, all the parameters whose typical low-values are quantitatively comparable with the minimum difference between consecutive eigenvalues of the perfect system should be considered *imperfections*, that is, should be mathematically treated as $\varepsilon$-ordered perturbations. Thus, paying due attention, an effective and uniformly valid asymptotically approximate eigensolution (the surface $\mathcal{S}_\epsilon$ in Figure 2) is expected in a large neighborhood $\mathcal{M}^\circ$ of $\boldsymbol{\mu}^\circ$, spanned by the $\epsilon$-amplitude $\boldsymbol{\mu}'$-vector.

According to these criteria, the $\epsilon$-order smallness attributed to the parameters in (37) has been based on physical considerations, as far as it accounts for the typical low-density distribution of the rings (small $\delta$), which usually possess a limited rotational-to-translational mass ratio (small $\chi$) and are interconnected by highly-slender ligaments (small $\varrho$). On the contrary, the $\mathcal{O}(1)$-order attributed – for instance – to the aspect ratio $\eta$ means that extreme geometric cases (strongly rectangular cells) require a different, specific treatment.

In the absence of $\mathbf{H}^\circ$-defectivity and assuming a regular (non-singular) perturbation $\epsilon\boldsymbol{\mu}'$, the exact eigensolution of the matrix $\mathbf{H}(\boldsymbol{\mu})$ can be approximated by a series function $\lambda(\epsilon)$ of integer $\epsilon$-powers [44]. For the generic eigenvalue it reads

$$\lambda(\epsilon) = \lambda^\circ + \sum_n \lambda^{(n)}\epsilon^n = \lambda^\circ + \lambda'\epsilon + \lambda''\epsilon^2 + \ldots + \lambda^{(n)}\epsilon^n + \ldots \tag{38}$$

where $n!\lambda^{(n)}$ can be regarded as the $n$-th $\epsilon$-derivative (for $\epsilon = 0$) of the exact but implicit eigenvalue function $F(\lambda, \epsilon) = 0$. Geometrically, it represents a directional derivative in the $\boldsymbol{\mu}'$-direction of the parameter space (evaluated in $\boldsymbol{\mu} = \boldsymbol{\mu}^\circ$), and is commonly defined as the $n$-th eigenvalue sensitivity, or *eigensensitivity*. Since the *ideal* $(\beta_i, \lambda)$*-plane* $\Pi^\circ$ containing the reference $\boldsymbol{\mu}^\circ$-point differs from the $\Pi$-plane, the curve $\lambda_\epsilon$ described by the approximate function $\lambda(\epsilon)$ is not required to cross (nor to be tangent to) the exact dispersion curve, in the general case (see Figure 2). The approximation accuracy is expected to decay with the $\epsilon$-measured distance from $\boldsymbol{\mu}^\circ$.

Once both the eigenvalue and parameter laws $\lambda(\epsilon)$ and $\boldsymbol{\mu}(\epsilon)$ have been assigned, the characteristic function can be expressed as a composite function $G(\epsilon) = F(\lambda(\epsilon), \boldsymbol{\mu}(\epsilon))$, which admits the Maclaurin expansion in $\epsilon$-power series

$$G(\epsilon) = G^\circ + \sum_n \frac{G^{(n)}}{n!}\epsilon^n = G^\circ + G'\epsilon + \frac{G''}{2}\epsilon^2 + \ldots + \frac{G^{(n)}}{n!}\epsilon^n + \ldots \tag{39}$$

where, since $\lambda^\circ$ belongs to the $\mathbf{H}^\circ$-eigenspectrum, the constant $G^\circ$ is certainly null. The generic higher-order expansion coefficient $G^{(n)}$ is the $n$-th $\epsilon$-derivative (for $\epsilon = 0$) of the function $G(\epsilon)$, which can be calculated applying the chain rule of differentiation for *multi*-variable composite functions.

To specify, each coefficient $G^{(n)}$ is a complete $n$-degree polynomial of all the *unknowns* up to the $n$-th eigensensitivity. With reference to the $\epsilon$-power, the lowest-order coefficients read

$$\epsilon^1: \quad G' = \lambda' F^{(1,0)} + C'_\mu \tag{40}$$

$$\epsilon^2: \quad G'' = 2\lambda'' F^{(1,0)} + (\lambda')^2 F^{(2,0)} + 2\lambda'(\boldsymbol{\mu}')^\top \mathbf{F}^{(1,1)} + C''_\mu \tag{41}$$

where the parameter-dependent *known* terms are

$$C'_\mu = (\boldsymbol{\mu}')^\top \mathbf{F}^{(0,1)}, \quad C''_\mu = 2(\boldsymbol{\mu}'')^\top \mathbf{F}^{(0,1)} + (\boldsymbol{\mu}')^\top \mathbf{F}^{(0,2)}\boldsymbol{\mu}' \tag{42}$$

while the synthetic notation $F^{(h,0)} = \partial^h_\lambda F(\lambda, \boldsymbol{\mu})$ for $h = 1, 2$ and $\mathbf{F}^{(h,k)} = \partial^h_\lambda \partial^k_{\boldsymbol{\mu}} F(\lambda, \boldsymbol{\mu})$ for $h, k = 0, 1, 2$ is adopted for the partial derivatives of the characteristic function.

As additional original contribution, the Scott-Faà di Bruno's formula [43] has been further manipulated to deal with the specific multi-parameter case, which requires the treatment of composite derivatives with multi-variable inner functions. Recurring to the multi-index notation, the $n$-th coefficient can be expressed by the recursive but explicit formula

$$\epsilon^n: \quad G^{(n)} = \sum_{\mathcal{S}(h,k)} \sum_{|p|=k} \frac{n!}{h+k} \mathbf{F}^{(h,|p|)} \boldsymbol{\eta}^{[n]}_{hp} \tag{43}$$

where, introducing the multi-index $p$ (with $|p| = k$), the partial derivatives for generic $h, k \in \mathbb{Z}^+$ can be expressed

$$\mathbf{F}^{(h,|p|)} = \partial^h_\lambda \partial^{|p|}_{\boldsymbol{\mu}} F(\lambda, \boldsymbol{\mu}) = \frac{\partial^h}{\partial\lambda^h} \frac{\partial^{|p|} F(\lambda, \mu_{p_1}, \ldots, \mu_{p_k})}{\partial\mu_{p_1} \ldots \partial\mu_{p_k}} \tag{44}$$

while the $(h,k)$-index set is again $\mathcal{S}(h,k) = (\,h, k \in [0, h+k = n])$. Defining $\ell$ the dimension of the $\boldsymbol{\mu}$-vector, the multi-parametric recursion is a non-differential, polynomial function

$$\boldsymbol{\eta}^{[n]}_{hp} = \sum_{i=1}^{\ell} \frac{1}{n\,\mu^\circ_i \lambda^\circ} \sum_{j=1}^{n} (j(h+k) + j - n)\lambda^{(j)} \mu^{(j)}_i \boldsymbol{\eta}^{[n-j]}_{hp} \tag{45}$$

and has to be initialized with $\boldsymbol{\eta}^{[0]}_{hp} = (\lambda^\circ)^h (\boldsymbol{\mu}^\circ)^p$. The brief notes in the AppendixB clarify how the specific nature of the power series $\lambda(\epsilon)$ and $\boldsymbol{\mu}(\epsilon)$ has been leveraged to obtain the result. A simpler closed form of the equation (43), expressed in terms of the vector variable $\boldsymbol{\nu}(\epsilon) = (\lambda(\epsilon), \boldsymbol{\mu}(\epsilon))$, is also presented. Of course, the single-parameter formula (31) can be extracted as particular case from the multi-parameter formula (43).

Similarly to the *SPPM*, the *MPPM* requires the approximate characteristic function to be satisfied by zeroing each $\epsilon^{(n)}$-order coefficient $G^{(n)}$. Thus, an ordered chain of $n$ perturbation equations is obtained, and the solution algorithm depends on the individual $m^\circ$-multiplicity of the generating solution represented by the twelve eigenvalues $\lambda^\circ$ satisfying the zeroth-order equation $F(\lambda^\circ, \boldsymbol{\mu}^\circ) = 0$. Again, two fundamental cases occur

- *Simple eigenvalue*: if $\lambda^\circ$ is a simple root ($m^\circ = 1$) for the equation $F(\lambda^\circ, \boldsymbol{\mu}^\circ) = 0$, then the coefficient $F^{(1,0)} \neq 0$. Hence, the $\epsilon^1$-order equation (40) is linear in the unknown $\lambda'$, the $\epsilon^2$-order equation (41) is linear in the unknown $\lambda''$ and so on. Therefore, the cascade solution (null if the numerator vanishes) for the lowest order eigensensitivities is

$$\epsilon^1: \quad \lambda' = -\frac{C'_\mu}{F^{(1,0)}} \tag{46}$$

$$\epsilon^2: \quad \lambda'' = -\frac{2\lambda'(\boldsymbol{\mu}')^\top \mathbf{F}^{(1,1)} + (\lambda')^2 F^{(2,0)} + C''_\mu}{2F^{(1,0)}} \tag{47}$$

  and, by extension, the $\epsilon^{(n)}$-order equation allows the determination of the $n$-th eigensensitivity.

- *Double (semi-simple) eigenvalue*: if $\lambda^\circ$ is a double root ($m^\circ = 2$) for the equation $F(\lambda^\circ, \boldsymbol{\mu}^\circ) = 0$, then the coefficient $F^{(1,0)} = 0$, but $F^{(2,0)} \neq 0$. Since $\lambda^\circ$ must be non-defective (semi-simple), it can be proved that $C'_\mu = 0$ [44]. Consequently, the $\epsilon^1$-order equation (40) is trivially satisfied, but leaves $\lambda'$ undetermined. Such an indetermination is cleared by the $\epsilon^2$-order equation (41), which is a quadratic in the $\lambda'$-unknown only (as the $\lambda''$-coefficient is null). Thus, the solution of the lowest order equations is

$$\epsilon^1: \quad \lambda' \quad \text{undetermined} \tag{48}$$

$$\epsilon^2: \quad \lambda'_\pm = -\frac{(\boldsymbol{\mu}')^\top \mathbf{F}^{(1,1)} \pm \sqrt{((\boldsymbol{\mu}')^\top \mathbf{F}^{(1,1)})^2 - C''_\mu F^{(2,0)}}}{F^{(2,0)}} \tag{49}$$

  where the eigensensitivity pair $\lambda'_\pm$ splits the double eigenvalue $\lambda^\circ$ in two simple eigenvalues $\lambda^\circ + \epsilon\lambda'_\pm + O(\epsilon^2)$. The occurrence of a vanishing discriminant is a particular case, in which the eigenvalue splitting is postponed to higher-orders. In the general case, instead, the higher eigensensitivities $\lambda_\pm^{(n)}$ are determined by linear $\epsilon^{(n+1)}$-order equations, solved for one or the other of the $\lambda'_\pm$-values.

Of course, more involved cases concerning higher eigenvalue multiplicity (say $m^\circ > 2$) can be encountered and must be approached with similar solution schemes. As general rule, apart nested degeneration cases, the first sensitivity $\lambda'$ of a $m^\circ$-tuple eigenvalue $\lambda^\circ$ is expected to remain undetermined up to the $m^\circ$-th order equation, which is pure in the $\lambda'$-unknown and supplies $m^\circ$ different roots, which split the multiple eigenvalue into a cluster of close eigenvalues separated by $\epsilon$-order differences.

When the *MPPM* is applied to the eigenproblem (26) for $\beta_i^\circ = 0$ (left bound of the $\mathcal{B}_i$-range), the generating solution includes cases of single, double and quadruple $\lambda^\circ$-eigenvalues. Table 4 reports the solution scheme required to achieve a fourth-order approximation $\lambda_i(\epsilon) = \lambda_i^\circ + \lambda'_i\epsilon + \lambda''_i\epsilon^2 + \lambda'''_i\epsilon^3 + \lambda''''_i\epsilon^4 + O(\epsilon^5)$ of all the twelve eigenvalues ($i = 1, ..., 12$, one for each row). As interesting remark, a quadruple null eigenvalue exists ($\lambda^\circ_{1..4} = 0$), which remains unsplitted at the $\epsilon^4$-order (due to the vanishing discriminant governing the splitting process). Hence, its fourth-order approximation requires the solution of the perturbation equations up to the $\epsilon^{10}$-order. The same solution scheme has been found to properly work for both the rectangular and the square cell. As complementary remark, the real-valued eigenvectors $\boldsymbol{\varphi}_a^\circ$ corresponding to the $\lambda^\circ$ eigenvalues possess purely local shapes, meaning that the ideal cell possesses purely translational or rotational modes (Figure 4). To complete the analysis, all the $\lambda^\circ$-eigenvalues for $\beta_i^\circ = \pi$ (right bound of the $\mathcal{B}_i$-range) have been verified to possess a double multiplicity.

A summarizing discussion about the differences and analogies between the *SPPM* and *MPPM* can briefly be outlined, as far as they are general tools for the Floquet-Bloch analysis of – potentially – a large variety of cellular materials. First, the two methods differ in the selection criterion of the unperturbed system, governed by the matrices $\mathbf{H}^*$ and $\mathbf{H}^\circ$, respectively, and employed as reference point to start the perturbation analysis. This difference generally translates into dissimilar generating

Table 3: *MPPM* asymptotic approximation: zeroth-order eigenvalue multiplicity and eigensensitivity solution scheme for $\beta_i^\circ = 0$.

| $m^\circ$ | $\epsilon^0$ | $\epsilon^1$ | $\epsilon^2$ | $\epsilon^3$ | $\epsilon^4$ | $\epsilon^5$ | $\epsilon^6$ | $\epsilon^7$ | $\epsilon^8$ | $\epsilon^9$ | $\epsilon^{10}$ |
|---|---|---|---|---|---|---|---|---|---|---|---|
| 4 | $\lambda^\circ_{1...4}$ | - | - | - | $\lambda'_{1...4}$ | - | - | - | $\lambda''_1$<br>$\lambda''_2$<br>$\lambda''_3$<br>$\lambda''_4$ | $\lambda'''_1$<br>$\lambda'''_2$<br>$\lambda'''_3$<br>$\lambda'''_4$ | $\lambda''''_1$<br>$\lambda''''_2$<br>$\lambda''''_3$<br>$\lambda''''_4$ |
| 2 | $\lambda^\circ_{5,6}$ | - | $\lambda'_{5,6}$ | - | $\lambda''_5$<br>$\lambda''_6$ | $\lambda'''_5$<br>$\lambda'''_6$ | $\lambda''''_5$<br>$\lambda''''_6$ | ...<br>... | ...<br>... | ...<br>... | ...<br>... |
| 2 | $\lambda^\circ_{7,8}$ | - | $\lambda'_{7,8}$ | - | $\lambda''_7$<br>$\lambda''_8$ | $\lambda'''_7$<br>$\lambda'''_8$ | $\lambda''''_7$<br>$\lambda''''_8$ | ...<br>... | ...<br>... | ...<br>... | ...<br>... |
| 1 | $\lambda^\circ_9$ | $\lambda'_9$ | $\lambda''_9$ | $\lambda'''_9$ | $\lambda''''_9$ | ... | ... | ... | ... | ... | ... |
| 1 | $\lambda^\circ_{10}$ | $\lambda'_{10}$ | $\lambda''_{10}$ | $\lambda'''_{10}$ | $\lambda''''_{10}$ | ... | ... | ... | ... | ... | ... |
| 1 | $\lambda^\circ_{11}$ | $\lambda'_{11}$ | $\lambda''_{11}$ | $\lambda'''_{11}$ | $\lambda''''_{11}$ | ... | ... | ... | ... | ... | ... |
| 1 | $\lambda^\circ_{12}$ | $\lambda'_{12}$ | $\lambda''_{12}$ | $\lambda'''_{12}$ | $\lambda''''_{12}$ | ... | ... | ... | ... | ... | ... |

Legend: “-” stands for undetermined, “...” means higher sensitivities solutions, requiring distinct solution schemes for the perturbation equations. This remark prevents considering the *SPPM* as a particular case of the *MPPM*, apart from some aspects of the mathematical formalism. In the practice, may also be impossible to systematically predict which method requires the shorter $F(\lambda, \boldsymbol{\mu})$-expansion to achieve the same approximation order. Moreover, the distance separating the starting points $\boldsymbol{\mu}^*$ and $\boldsymbol{\mu}^\circ$ in the parameter space implies a mismatch in the local approximation accuracy even for the same parameter set $\boldsymbol{\mu}$, depending on the different perturbation amplitudes, measured by the $\xi$ and $\epsilon$-parameters, respectively. Therefore, a global comparison between the methods tends to be essentially worthless in quantitative terms. Nonetheless, qualitative recurrences in the solution schemes can be recognized, depending on the eigenvalue multiplicity of the unperturbed matrices. Such schemes are featured by some formal similarities, as evident from the comparison of the lowest-order perturbation equations (33)-(34) and (46)-(47), which differ only by the coefficients of the unknowns. As minor remark, the approximate solution given by the *MPPM* tends to be less cumbersome, since all the coefficients (including the known terms $C_\mu^{(n)}$) are polynomial functions of the multiparameter perturbation. As minor technical difference, the convenient form $\lambda(\beta_i)$ of the approximate solution can be assessed (i) in the *SPPM* by simply employing the relation $\xi = \beta_i - \beta_i^*$ to replace the local abscissa; (ii) in the *MPPM* by inverting all the parameter ordering relations (*backward rescaling*)

$$\beta_i' = \frac{\beta_i - \beta_i^\circ}{\epsilon}, \qquad \delta' = \frac{\delta}{\epsilon}, \qquad \varrho' = \frac{\varrho}{\epsilon}, \qquad \chi' = \frac{\chi}{\epsilon} \tag{50}$$

to replace the multiparameter parameter perturbation (the $\epsilon$-independent inverse relations $\eta^\circ = \eta$ and $\omega_c^\circ = \omega_c$ are also understood). Finally, the substitution of the inverse relations (50) into the approximation function (38) allows the complete simplification (*reabsorption*) of the auxiliary $\epsilon$-parameter. Of course, the reabsorption procedure does not alter the approximation order.

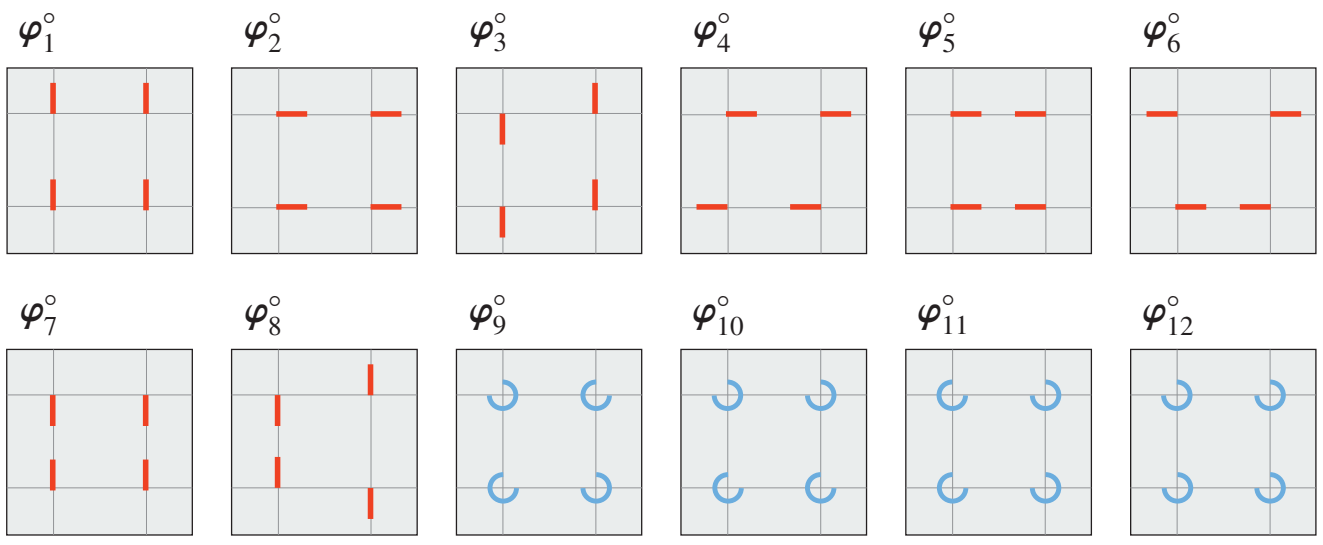

Figure 4: *MPPM* asymptotic approximation: purely translational (red) or rotational (blue) local modes (subscript $a$ omitted) of the ideal cell.

### 3.3. Results and discussion

Employing standard techniques of solution continuation for the characteristic equation governing the eigenproblem (26), the Floquet-Bloch spectrum has been analysed. Focus has been made on the horizontal direction of wave propagation, by carrying out the twelve $\omega$-frequency loci under variation of the $\beta_1$-wavenumber in the irreducible range ($\beta_1 \in \mathcal{B}_1$). A particular set of nondimensional mechanical parameters has been selected, coherently with the technical and constructional requirements of massive rings coupled with slender ligaments ($\chi^2 = 1/81, \delta = 1/10, \varrho^2 = 1/100$). The exact dispersion curves for the rectangular ($\eta = 3/2$) and square cell ($\eta = 1$) are reported in Figure 5 (red lines) and Figure 6 (blue lines), respectively. Of course, for symmetry reasons, the results concerning the square cell can be indifferently referred to the horizontal ($\beta_1$) or vertical ($\beta_2$) wave directions, whereas supplementary results concerning the rectangular cell with different aspect ratios ($\eta = 2, \eta = 1/2$) have been presented in [40].

As major qualitative remark, the variation ranges of the low-frequencies (namely $\omega_1$-$\omega_8$) and high-frequencies (namely $\omega_9$-$\omega_{12}$) tend to be well-separated and mainly dominated by translational and rotational modes, respectively. Parametric analyses, here not reported for the sake of synthesis, show how the *translational frequencies* (rigorously, the frequencies of translational modes) undergo only minimal qualitative changes if the parameter set ($\varrho^2, \chi^2, \delta$) vary in the technically-relevant range. On the contrary, the *rotational frequencies* (rigorously, the frequencies of rotational modes) strongly depend on the $\delta$ and $\chi$ parameters, with an approximately linear law of direct and inverse proportionality, respectively. Finally, growing $\varrho^2$-values let all the rotational frequencies become increasingly close to each other, fixed the highest frequency root [40].

As technically-relevant result, the rectangular geometry does not present low-frequency band gaps among the translational curves, and extensive parametric analyses have confirmed this finding along both the orthogonal propagation directions for a generic parameter set. The absence of band-gaps is mainly due to the structural property of the governing matrix, which forces the curve ends to coincide in pairs at each boundary of the $\mathcal{B}_1$-range. The possibility of band-gaps is also limited by the frequent occurrence of crossing points in the mid $\mathcal{B}_1$-range, involving curve pairs of some consecutive low-frequencies (2nd-3rd or 4th-5th, for instance). On the contrary, two well-distinct band-gaps may exist in the high frequency range, falling between the roots of translational-rotational curves (gray region $BG_I$) or rotational-rotational curves (gray region $BG_{II}$). The latter is certainly a partial band-gap, since it can be verified to disappear along the short-side cell direction (for certain $\beta_2$-wavenumbers). The gap amplitudes can be verified to increase proportionally to an increment of the $\delta$-values (or $\varrho^2$-values, taking fixed the highest frequency $\omega_{12}$) or, alternatively, to a decrement of the $\chi^2$-values. The square shape closes the band gap $BG_{II}$ at higher frequency, whereas a small-amplitude partial band-gap opens at low-frequencies along the diagonal direction of wave propagation [40], spanned by the combination wavenumber $\beta_{12} = (\beta_1^2 + \beta_2^2)^{1/2}$ with $\beta_1 = \beta_2$, and corresponding to the direction of minimum auxeticity of the material [26].

Taking the left $\mathcal{B}_1$-boundary as reference point ($\beta_1^* = \beta_1^\circ = 0$), the asymptotic perturbation approximations of the twelve eigenvalues are reported as explicit functions of the parameters in Table 4 for the *SPPM* (before the replacement of the $\xi$-parameter) and Table 5 for the *MPPM* (before the reabsorption of the $\epsilon$-parameter). Independently of the approximation method, a remarkable property of the asymptotic perturbation solutions is the absence of all the series terms with odd-power order (say $\xi^n$ or $\epsilon^n$, with $n = 1, 3$), due to the null value determined for by the corresponding sensitivities ($\dot{\lambda}_i = \dddot{\lambda}_i = 0$, or $\lambda_i' = \lambda_i''' = 0$). Such remark can be mathematically justified by the inherent symmetry of all the spectrum branches with respect to the axis $\beta_i = 0$. This property reflects into the evenness of all the analytical functions which describe the dispersion curves by adopting the null value as origin of the independent $\beta_i$-variable, including – as particular case – the two perturbation expansions which employ $\beta_1^* = \beta_1^\circ = 0$ as unperturbed point. For the sake of completeness, it can be remarked that the spectrum branches are anti-symmetric with respect to the axis $\beta_i = \pi$. All the $\mathcal{Y}$- and $\mathcal{J}$-coefficients multiplying the perturbation variables $\xi$ and $\epsilon$ can be verified to be $O(1)$, according to the different parameter ordering, and are reported in the Appendix. The corresponding approximate dispersion curves are represented by the circles (*SPPM*) and dots (*MPPM*) in Figures 5 and 6. Despite the relatively-low (fourth-order) approximation, the asymptotic results turn out to closely match the exact solution throughout a large range of the varying parameters. In particular, the approximate solutions well-fit the twelve dispersion curves in almost the whole $\mathcal{B}_1$-range, and (somehow unexpectedly) not only in the closest neighborhood of the reference value. Indeed, a satisfying effectiveness of the asymptotic approximations generally persists up to the opposite boundary, expect for minor mismatches in a few dispersion curves, no matter if the reference $\beta_1$-value is fixed at the left $\mathcal{B}_1$-boundary (*left approximations*, in Figures 5a and 6a) or right $\mathcal{B}_1$-boundary (*right approximations*, in Figures 5b and 6b).

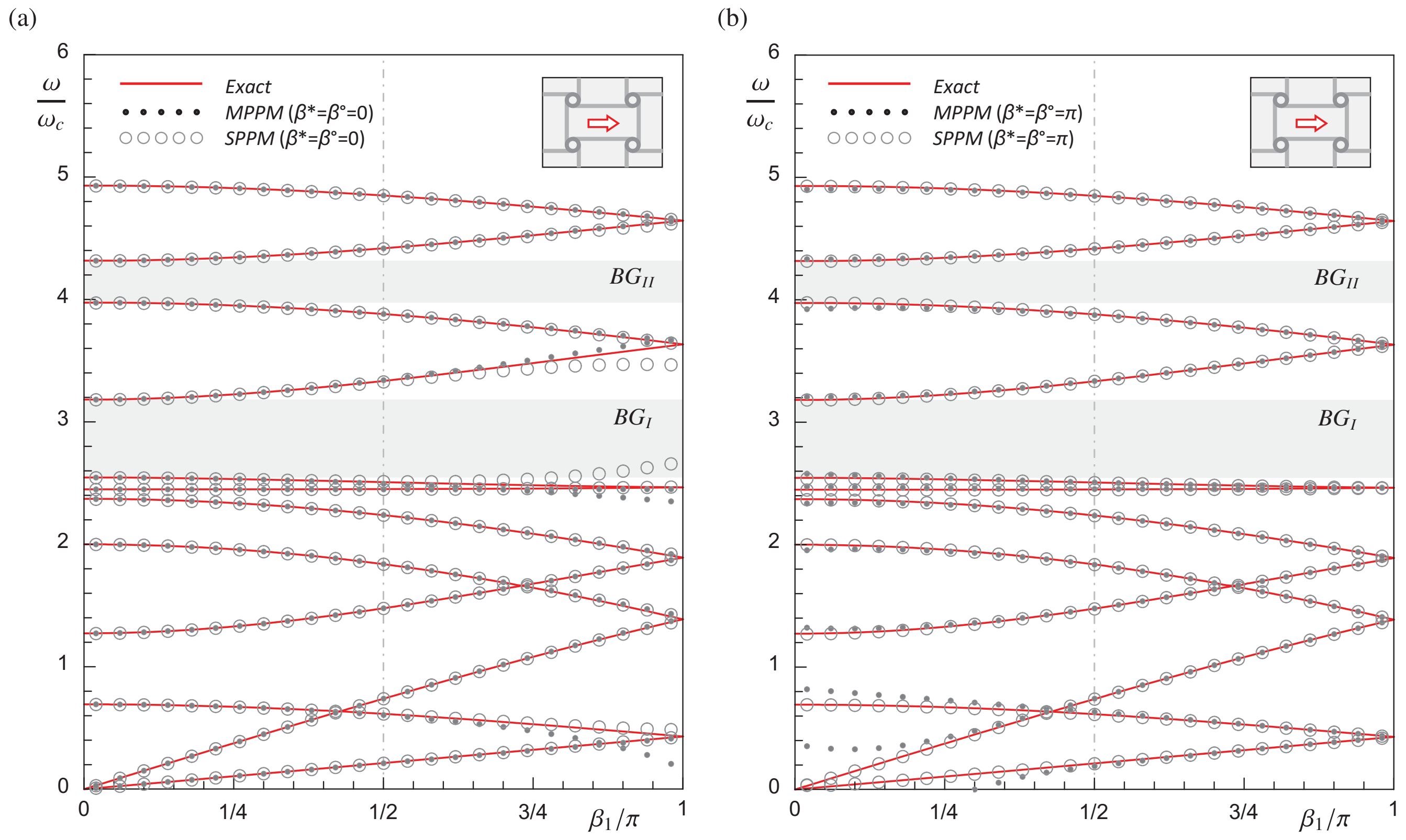

Figure 5: Floquet-Bloch spectrum of the rectangular cell ($\eta = 3/2$) in the $\beta_1$-direction: exact, *MPPM*- and *SPPM*-approximate frequency loci for (a) reference $\beta_1^* = \beta_1^\circ = 0$, (b) reference $\beta_1^* = \beta_1^\circ = \pi$.

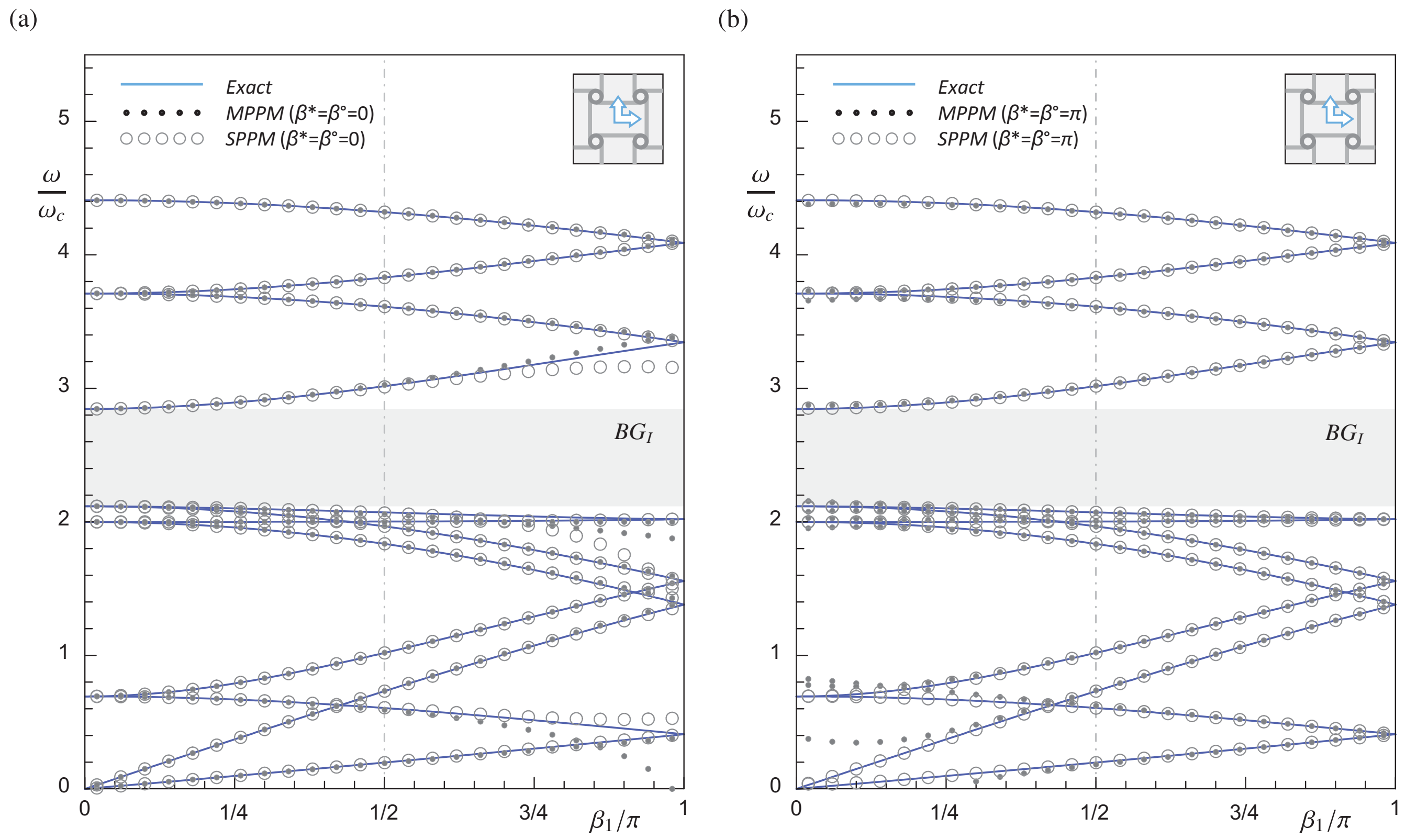

Figure 6: Floquet-Bloch spectrum of the square cell ($\eta = 1$) in the $\beta_1$-direction: exact, *MPPM*- and *SPPM*-approximate frequency loci for (a) reference $\beta_1^* = \beta_1^\circ = 0$, (b) reference $\beta_1^* = \beta_1^\circ = \pi$.

Table 4: *SPPM* asymptotic approximation: zeroth-order eigenvalue multiplicity and eigensensitivity solution scheme for $\beta_1^* = 0$.

| $\lambda_i(\xi)$ |
|---|
| $\lambda_1(\xi) = \lambda_1^* + \mathcal{Y}_{12}\xi^2 + \mathcal{Y}_{14}\xi^4 + O(\xi^5)$ |
| $\lambda_2(\xi) = \lambda_2^* + \mathcal{Y}_{22}\xi^2 + \mathcal{Y}_{24}\xi^4 + O(\xi^5)$ |
| $\lambda_3(\xi) = \lambda_3^* + \mathcal{Y}_{32}\xi^2 + \mathcal{Y}_{34}\xi^4 + O(\xi^5)$ |
| $\lambda_4(\xi) = \lambda_4^* + \mathcal{Y}_{42}\xi^2 + \mathcal{Y}_{44}\xi^4 + O(\xi^5)$ |
| $\lambda_5(\xi) = \lambda_5^* + \mathcal{Y}_{52}\xi^2 + \mathcal{Y}_{54}\xi^4 + O(\xi^5)$ |
| $\lambda_6(\xi) = \lambda_6^* + \mathcal{Y}_{62}\xi^2 + \mathcal{Y}_{64}\xi^4 + O(\xi^5)$ |
| $\lambda_7(\xi) = \lambda_7^* + \mathcal{Y}_{72}\xi^2 + \mathcal{Y}_{74}\xi^4 + O(\xi^5)$ |
| $\lambda_8(\xi) = \lambda_8^* + \mathcal{Y}_{82}\xi^2 + \mathcal{Y}_{84}\xi^4 + O(\xi^5)$ |
| $\lambda_9(\xi) = \lambda_9^* + \mathcal{Y}_{92}\xi^2 + \mathcal{Y}_{94}\xi^4 + O(\xi^5)$ |
| $\lambda_{10}(\xi) = \lambda_{10}^* + \mathcal{Y}_{102}\xi^2 + \mathcal{Y}_{104}\xi^4 + O(\xi^5)$ |
| $\lambda_{11}(\xi) = \lambda_{11}^* + \mathcal{Y}_{112}\xi^2 + \mathcal{Y}_{114}\xi^4 + O(\xi^5)$ |
| $\lambda_{12}(\xi) = \lambda_{12}^* + \mathcal{Y}_{122}\xi^2 + \mathcal{Y}_{124}\xi^4 + O(\xi^5)$ |

Note: $\xi = \beta_1$ can be understood as far as $\beta_1^* = 0$.

Table 5: *MPPM* asymptotic approximation: zeroth-order eigenvalue multiplicity and eigensensitivity solution scheme for $\beta_1^\circ = 0$.

| $\lambda_i(\epsilon)$ |
|---|
| $\lambda_1(\epsilon) = \lambda_1^\circ + \beta_1^2\varrho^2\mathcal{J}_{14}\epsilon^4 + O(\epsilon^5)$ |
| $\lambda_2(\epsilon) = \lambda_2^\circ + \beta_1^2\mathcal{J}_{22}\epsilon^2 + \beta_1^4\mathcal{J}_{24}\epsilon^4 + O(\epsilon^5)$ |
| $\lambda_3(\epsilon) = \lambda_3^\circ + \varrho^2\mathcal{J}_{32}\epsilon^2 + \beta_1^2\varrho^2\mathcal{J}_{34}\epsilon^4 + O(\epsilon^5)$ |
| $\lambda_4(\epsilon) = \lambda_4^\circ + \varrho^2\mathcal{J}_{42}\epsilon^2 + \left(\beta_1^4\mathcal{J}_{44}^\beta + \beta_1^2\varrho^2\mathcal{J}_{44}^\varrho\right)\epsilon^4 + O(\epsilon^5)$ |
| $\lambda_5(\epsilon) = \lambda_5^\circ + \beta_1^2\mathcal{J}_{52}\epsilon^2 + \left(\beta_1^4\mathcal{J}_{64}^\beta + \beta_1^2\delta^2\mathcal{J}_{64}^\delta\right)\epsilon^4 + O(\epsilon^5)$ |
| $\lambda_6(\epsilon) = \lambda_6^\circ + \left(\beta_1^2\mathcal{J}_{62}^\beta + \varrho^2\mathcal{J}_{62}^\varrho\right)\epsilon^2 + \left(\beta_1^4\mathcal{J}_{64}^\beta + \beta_1^2\delta^2\mathcal{J}_{64}^\delta\right)\epsilon^4 + O(\epsilon^5)$ |
| $\lambda_7(\epsilon) = \lambda_7^\circ + \beta_1^2\varrho^2\mathcal{J}_{74}\epsilon^4$ |
| $\lambda_8(\epsilon) = \lambda_8^\circ + \varrho^2\mathcal{J}_{82}\epsilon^2 + \beta_1^2\varrho^2\mathcal{J}_{84}\epsilon^4 + O(\epsilon^5)$ |
| $\lambda_9(\epsilon) = \lambda_9^\circ + \beta_1^2\mathcal{J}_{92}\epsilon^2 + \left(\beta_1^4\mathcal{J}_{94}^\beta + \varrho^2\beta_1^2\mathcal{J}_{94}^\varrho + \delta^2\beta_1^2\mathcal{J}_{94}^\delta\right)\epsilon^4 + O(\epsilon^5)$ |
| $\lambda_{10}(\epsilon) = \lambda_{10}^\circ + \beta_1^2\mathcal{J}_{102}\epsilon^2 + \left(\beta_1^4\mathcal{J}_{104}^\beta + \varrho^2\beta_1^2\mathcal{J}_{104}^\varrho + \delta^2\beta_1^2\mathcal{J}_{104}^\delta\right)\epsilon^4 + O(\epsilon^5)$ |
| $\lambda_{11}(\epsilon) = \lambda_{11}^\circ + \beta_1^2\mathcal{J}_{112}\epsilon^2 + \left(\beta_1^4\mathcal{J}_{114}^\beta + \varrho^2\beta_1^2\mathcal{J}_{114}^\varrho + \delta^2\beta_1^2\mathcal{J}_{114}^\delta\right)\epsilon^4 + O(\epsilon^5)$ |
| $\lambda_{12}(\epsilon) = \lambda_{12}^\circ + \beta_1^2\mathcal{J}_{122}\epsilon^2 + \left(\beta_1^4\mathcal{J}_{124}^\beta + \varrho^2\beta_1^2\mathcal{J}_{124}^\varrho + \delta^2\beta_1^2\mathcal{J}_{124}^\delta\right)\epsilon^4 + O(\epsilon^5)$ |

Note: apex is omitted for the parameters $\beta_1$, $\delta$ and $\varrho$

Although marginal, the mismatches with respect to the exact solution may deserve some further attention, in so far as they can strongly be reduced by properly combining the companion left and right approximations, conventionally referred to as $\lambda_i^-$ and $\lambda_i^+$ in the following. Among the others, a suited possibility is to adopt the linear and weighted combination

$$\lambda_i^\mp = w^-\lambda_i^- + w^+\lambda_i^+ \tag{51}$$

where $w^-$ and $w^-$ stand for the weighting $\beta_1$-dependent functions, required to be complementary to each other ($w^+ = 1 - w^-$) and attain null boundary values ($w^+ = 0$ and $w^- = 0$ at the left and right $\mathcal{B}_1$-boundary, respectively). Consistently, the weight of each approximation function is maximum in the closest proximity of its own reference point, whereas it decays up to vanish in the closeness of the reference point of the companion function. It is worth remarking that this combination differs from the matched composition of inner and outer asymptotic expansions (often employed in the presence of singularities), as far as there is no need for the companion functions to attain the same value (matching) in a certain limit point of their domain.

To the specific purposes of the present study, each combination $\lambda_i^\mp$ (for $i = 1...12$) has been build by adopting the highly-adaptable pair of transcendental weighting functions

$$w^- = \frac{1}{2} - \frac{\tanh\left(\gamma\left(2\beta - \pi\right)\right)}{2\tanh\left(\gamma\pi\right)}, \qquad w^+ = \frac{1}{2} + \frac{\tanh\left(\gamma\left(2\beta - \pi\right)\right)}{2\tanh\left(\gamma\pi\right)} \tag{52}$$

where the $\gamma$-parameter governs the higher (small $\gamma$-values) or lower (large $\gamma$-values) smoothness of the transition from unity to zero across the mid $\mathcal{B}_1$-range (centered in $\beta_1 = \pi/2$). The weighted combination $\lambda_i^*$ is reported in Figures 7 for all the twelve dispersion curves (for $\gamma = 2$). A satisfying agreement among the exact and approximate solutions in the full $\beta$-range can be appreciated for both the rectangular and the square cell, for both the *SPPM* and the *MPPM* approximations. As minor remark, this result is mostly inherent to the high-quality of the asymptotic perturbation approximations, and can be proved to persist for different choices of the weighting functions.

As characterizing aspect of theoretical and applied interest, the two dispersion curves related to the lowest frequencies ($i = 1, 2$), featured by null roots, represent the acoustic branches of the Floquet-Bloch spectrum [31, 38, 39], with nondimensional

- phase velocity $\quad c_{pi} = \omega_i/\beta$
- group velocity $\quad c_{gi} = \partial\omega_i/\partial\beta$

which are related to the respective dimensional counterparts $C_{pi}$ and $C_{gi}$ by the relations $C_{pi} = 2c_{pi}\Omega_r L_h$ and $C_{gi} = 2c_{gi}\Omega_r L_h$.

Together, the quantities $c_{pi}$ and $c_{gi}$ comprehensively characterize the propagation velocity of the acoustic waves with shear form (for $i = 1$, with $\pm\frac{1}{2}\pi$ polarization angle like $\boldsymbol{\varphi}_1^\circ$ in Figure 4) or compression form (for $i = 2$, with $0, \pi$ polarization angle like $\boldsymbol{\varphi}_2^\circ$ in Figure 4). The exact velocity curves versus the varying $\beta_1$-parameter are reported in Figure 8 for the rectangular ($\eta = 3/2$, red lines in Figure 8a) and square cell ($\eta = 1$, blue lines in Figure 8b). It can be remarked that the (lower) velocities of the shear waves ($c_{p1}, c_{g1}$) are almost independent of the $\beta_1$-parameter in the whole $\mathcal{B}_1$-range, whereas the (higher) velocities of the compression waves ($c_{p2}, c_{g2}$) nonlinearly but monotonically decrease for increasing $\beta_1$-parameters. The relation nonlinearity can be verified to grow up for decreasing ligament slenderness $\varrho^2$. Only minor quantitative effects can be attributed to different aspect ratios $\eta$, the most important being a small but evident decrement of the shear wave velocities in the square cell. As general remark, the phase and group velocities of each wave are quite similar to each other, meaning that each harmonic component propagates with almost the same velocity of the wave envelope. Further parametric analyses have shown that the shear wave velocity may be significantly sensitive to variations of the ligament slenderness $\varrho^2$, with faster waves related to ligaments with lower slenderness (higher $\varrho^2$-values).

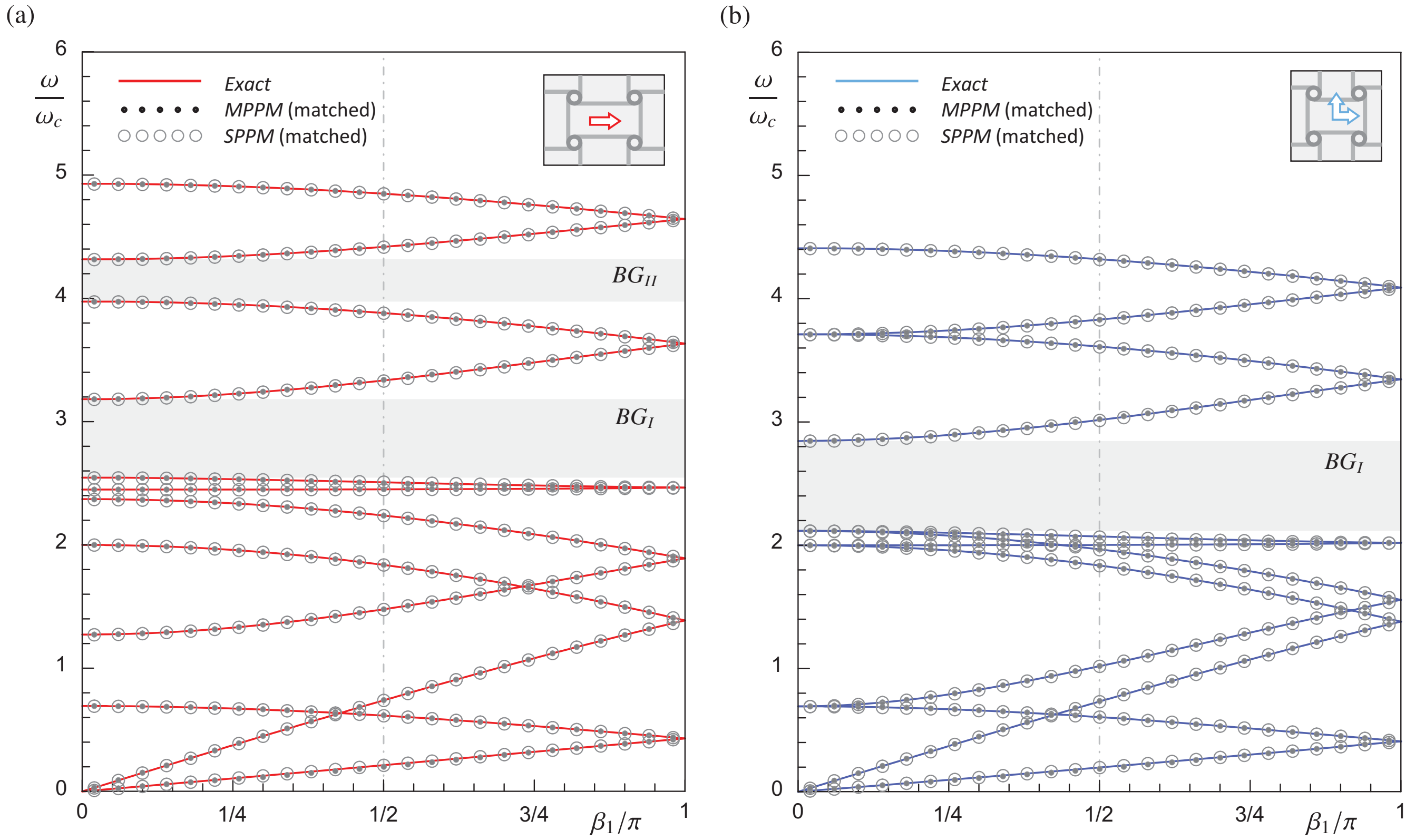

Figure 7: Floquet-Bloch spectrum in the $\beta_1$-direction: exact versus matched *MPPM*- and *SPPM*-approximate frequency loci for (a) rectangular cell ($\eta = 3/2$), (b) square cell ($\eta = 1$).

By virtue of the asymptotic perturbation solutions, all the wave velocities can be determined as explicit, although approximate, functions of the parameters. According to the *SPPM* scheme and consistently with the fourth-order (left) approximation of the eigenvalues, the phase and group velocities read

$$c_{p1} = \omega_c \sqrt{\mathcal{Y}_{12} + \beta^2 \mathcal{Y}_{14}}, \qquad c_{p2} = \omega_c \sqrt{\mathcal{Y}_{22} + \beta^2 \mathcal{Y}_{24}} \tag{53}$$

$$c_{g1} = \omega_c \frac{\mathcal{Y}_{12} + 2\beta^2 \mathcal{Y}_{14}}{\sqrt{\mathcal{Y}_{12} + \beta^2 \mathcal{Y}_{14}}}, \qquad c_{g2} = \omega_c \frac{\mathcal{Y}_{22} + 2\beta^2 \mathcal{Y}_{24}}{\sqrt{\mathcal{Y}_{22} + \beta^2 \mathcal{Y}_{24}}} \tag{54}$$

whereas according to the *MPPM* scheme they read

$$c_{p1} = \varrho \omega_c \sqrt{\mathcal{J}_{14}}, \qquad c_{p2} = \omega_c \sqrt{\mathcal{J}_{22} + \beta^2 \mathcal{J}_{24}} \tag{55}$$

$$c_{g1} = \varrho \omega_c \sqrt{\mathcal{J}_{14}}, \qquad c_{g2} = \omega_c \frac{\mathcal{J}_{22} + 2\beta^2 \mathcal{J}_{24}}{\sqrt{\mathcal{J}_{22} + \beta^2 \mathcal{J}_{24}}} \tag{56}$$

where all the coefficients $\mathcal{Y}$ and $\mathcal{J}$ are $\beta$-independent $O(1)$-terms, as reported in the Appendix. In order to exemplify the synthetic descriptive potential of the perturbation-based results, suited for technical application and design purposes, the dimensional form of the shear wave velocities read

$$C_{p1} = C_{g1} = 2\sqrt{\frac{EI}{M(L_h + L_v)}} \tag{57}$$

according to the *MPPM* scheme. The *SPPM* scheme gives the same result if the fourth order is neglected.

The asymptotic perturbation approximations of the wave velocities are represented by the circles (*SPPM*) and dots (*MPPM*) in Figures 8. The high accuracy of the approximation can be appreciated in almost the whole $\mathcal{B}_1$-range, since the asymptotic results well-fit the exact velocities $c_{pi}$ and $c_{gi}$, apart for minimal differences occurring in the right $\mathcal{B}_1$-part, the farthest from the reference point. However, this minor loss of accuracy can be deemed not-negligible only for the high group velocities $c_{g2}$ of the compression wave. In particular, the *SPPM* and *MPPM* approximations are found to slightly under-estimate and over-estimate the exact $c_{g2}$-values, respectively, no matter the aspect ratio. Finally, it may be worth noting that the shear-wave velocities are constant in the $\mathcal{B}_1$-range, that is, $\beta_1$-independent, according to the *MPPM* approximations.

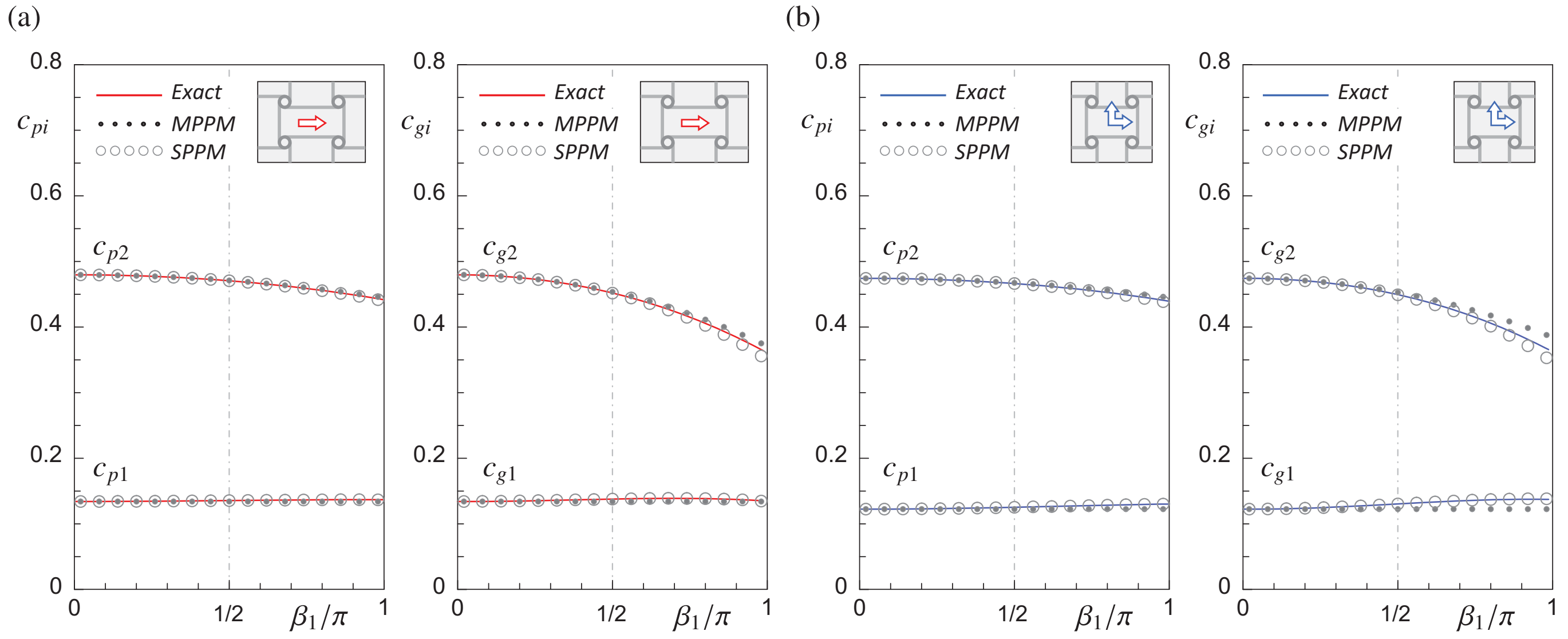

Figure 8: Exact versus *MPPM*- or *SPPM*-approximate phase velocity $c$ and group velocity $c_g$ in the $\beta_1$-direction for the acoustic wave branches of the Floquet-Bloch spectrum: (a) rectangular cell ($\eta = 3/2$), (b) square cell ($\eta = 1$)

## 4. Conclusions

The wave propagation features offered by a novel class of smart cellular materials, characterized by marked auxetic properties, have been analyzed. The periodic cell microstructure is composed by a regular pattern of equispaced rings connected by tangent ligaments. The material auxeticity is provided by the peculiar anti-tetrachiral symmetry of the structural geometry, which activates pervasive rolling-up mechanisms of the ring-ligament system. Adopting a rigid body assumption for the stiff rings and the unshearable beam theory for the flexible ligaments, a linear parametric model with lumped mass and distributed elasticity has been formulated to govern the free dynamic response of the elementary cell. Employing a static condensation of the passive degrees-of-freedom, a general procedure has been adopted to impose the quasi-periodicity conditions of free wave propagation in the low-dimension space of the twelve active degrees-of-freedom. Based on the Floquet-Bloch theory, the solution of the resulting eigenproblem (with real-valued eigenvalues, complex-valued eigenvectors) has furnished the dispersion curves and polarization modes of the propagation waves traveling along the periodic material.

First, the wavenumber-dependent eigensolution has been carried out by traditional techniques of numerical continuation. Parametric analyses have disclosed a rich Floquet-Bloch spectrum, in which translational and rotational modes tend to dwell in the low- and high-frequency ranges, respectively. The persistent absence of low-frequency band-gaps has been verified for different parameter sets, as natural consequence of the spectrum density, accompanied by the frequent occurrence of curve cross-overs. The effect of the main mechanical parameters on the dispersion curves has been discussed, with focus on the frequency modification and inter-frequency band amplification under variation of the cell aspect ratio, the ligament slenderness and the ring density.

Second, asymptotic perturbation techniques have been employed to attack the eigenproblem, in order to analytically assess the eigensolution as an explicit, although asymptotically approximate, functions of the structural parameters. Depending on the dimension of the perturbation vector, a *single-parameter* and a *multi-parameter* method have been distinguished and then separately developed, as far as they may require a different choice of the unperturbed (reference) point in the parameter space. The respective solution algorithms, based on a hierarchy of perturbation equations, have been defined in a systematic general form for the lowest approximation orders, and finally sketched out for the generic higher order, up to the desired approximation. Both the perturbation schemes have been successfully applied (up to the forth-order) to the particular problem under investigation, furnishing explicit parametric functions of the eigenvalues. Their comparison with the exact (numerical) solutions has shown a satisfying approximation accuracy over a wide range of the perturbation parameters, not limited to the closest neighborhood of the unperturbed values. Minor accuracy losses have been effectively removed by virtue of a suited weighted combination of approximate solutions with different reference points.

Finally, the phase and group velocities of the two acoustic branches, corresponding to shear and compression waves in the low-frequency range, have been determined. The asymptotic perturbation eigensolution has allowed the parametric assessment of the wave velocities, which has been verified to well-fit the exact numerical values. A simple but asymptotically consistent formula for the shear wave velocities has been given, in order to exemplify the synthetic descriptive potential of the perturbation-based results, suited for technical application and design purposes.

## AppendixA. Structural matrices

If the non-null 4-by-4 submatrices of the stiffness matrix $\mathbf{K}_{aa}$ in the equation (5) are expressed in the component form

$$\mathbf{K}_{uu} = \left[K^{uu}_{ij}\right], \quad \mathbf{K}_{vv} = \left[K^{vv}_{ij}\right], \quad \mathbf{K}_{\phi\phi} = \left[K^{\phi\phi}_{ij}\right], \quad \mathbf{K}_{u\phi} = \left[K^{u\phi}_{ij}\right], \quad \mathbf{K}_{v\phi} = \left[K^{v\phi}_{ij}\right], \tag{A.1}$$

for $i, j = 1...4$, their non-null components read

$$\begin{aligned}
&K^{uu}_{11} = K^{uu}_{22} = K^{uu}_{33} = K^{uu}_{44} = 3(1+36\eta^3\varrho^2)\\
&K^{uu}_{12} = K^{uu}_{21} = K^{uu}_{34} = K^{uu}_{43} = -1\\
&K^{uu}_{13} = K^{uu}_{31} = K^{uu}_{24} = K^{uu}_{42} = -12\eta^3\varrho^2\\
&K^{vv}_{11} = K^{vv}_{22} = K^{vv}_{33} = K^{vv}_{44} = 3(\eta+36\varrho^2)\\
&K^{vv}_{12} = K^{vv}_{21} = K^{vv}_{34} = K^{vv}_{43} = -12\varrho^2\\
&K^{vv}_{13} = K^{vv}_{31} = K^{vv}_{24} = K^{vv}_{42} = -\eta\\
&K^{\phi\phi}_{11} = K^{\phi\phi}_{22} = K^{\phi\phi}_{33} = K^{\phi\phi}_{44} = \tfrac{3}{4}(\eta+1)(\delta^2+16\varrho^2)\\
&K^{\phi\phi}_{12} = K^{\phi\phi}_{21} = K^{\phi\phi}_{34} = K^{\phi\phi}_{43} = \tfrac{1}{4}(8\varrho^2-\delta^2)\\
&K^{\phi\phi}_{13} = K^{\phi\phi}_{31} = K^{\phi\phi}_{24} = K^{\phi\phi}_{42} = -\tfrac{1}{4}\eta(\delta^2-8\varrho^2)\\
&K^{u\phi}_{11} = K^{u\phi}_{22} = -K^{u\phi}_{33} = -K^{u\phi}_{44} = \tfrac{1}{2}(\delta+36\eta^2\varrho^2)\\
&K^{u\phi}_{12} = K^{u\phi}_{21} = -K^{u\phi}_{34} = -K^{u\phi}_{43} = \tfrac{1}{2}\delta\\
&K^{u\phi}_{13} = -K^{u\phi}_{31} = K^{u\phi}_{24} = -K^{u\phi}_{42} = -6\eta^2\varrho^2\\
&K^{v\phi}_{11} = -K^{v\phi}_{22} = K^{v\phi}_{33} = -K^{v\phi}_{44} = \tfrac{1}{2}(\delta-36\varrho^2)\\
&K^{v\phi}_{12} = -K^{v\phi}_{21} = K^{v\phi}_{34} = -K^{v\phi}_{43} = 6\varrho^2\\
&K^{v\phi}_{13} = K^{v\phi}_{31} = -K^{v\phi}_{24} = -K^{v\phi}_{42} = \tfrac{1}{2}\eta\delta
\end{aligned} \tag{A.2}$$

Consistently with the decomposition of the matrix $\mathbf{K}_{aa}$, if the passive displacement vector is conveniently sorted and decomposed as $\mathbf{q}_p = (\mathbf{u}_p, \mathbf{v}_p, \boldsymbol{\phi}_p)$, with subvectors $\mathbf{u}_p = (u_5, ..., u_{12})$, $\mathbf{v}_p = (v_5, ..., v_{12})$ and $\boldsymbol{\phi}_p = (\phi_5, ..., \phi_{12})$, the global coupling matrix $\mathbf{K}_{ap}$ and the passive stiffness matrix $\mathbf{K}_{pp}$ can be expressed

$$\mathbf{K}_{ap} = \begin{pmatrix} \mathbf{C}_{uu} & \mathbf{O} & \mathbf{C}_{u\phi} \\ \mathbf{O} & \mathbf{C}_{vv} & \mathbf{C}_{v\phi} \\ \mathbf{C}_{\phi u} & \mathbf{C}_{\phi v} & \mathbf{C}_{\phi\phi} \end{pmatrix}, \quad \mathbf{K}_{pp} = \begin{pmatrix} \mathbf{P}_{uu} & \mathbf{O} & \mathbf{P}_{u\phi} \\ \mathbf{O} & \mathbf{P}_{vv} & \mathbf{P}_{v\phi} \\ \mathbf{P}_{\phi u} & \mathbf{P}_{\phi v} & \mathbf{P}_{\phi\phi} \end{pmatrix} \tag{A.3}$$

where $\mathbf{P}_{\phi u} = \mathbf{P}^\top_{u\phi}$ and $\mathbf{P}_{\phi v} = \mathbf{P}^\top_{v\phi}$ for the sake of symmetry. Therefore, if the non-null 4-by-8 submatrices (in $\mathbf{K}_{ap}$) and the 8-by-8 submatrices (in $\mathbf{K}_{pp}$) are expressed as

$$\mathbf{C}_{uu} = \left[C^{uu}_{ij}\right], \quad \mathbf{C}_{vv} = \left[C^{vv}_{ij}\right], \quad \mathbf{C}_{\phi\phi} = \left[C^{\phi\phi}_{ij}\right], \quad \mathbf{C}_{u\phi} = \left[C^{u\phi}_{ij}\right], \quad \mathbf{C}_{v\phi} = \left[C^{v\phi}_{ij}\right], \tag{A.4}$$

$$\mathbf{P}_{uu} = \left[P^{uu}_{jh}\right], \quad \mathbf{P}_{vv} = \left[P^{vv}_{jh}\right], \quad \mathbf{P}_{\phi\phi} = \left[P^{\phi\phi}_{jh}\right], \quad \mathbf{P}_{u\phi} = \left[P^{u\phi}_{jh}\right], \quad \mathbf{P}_{v\phi} = \left[P^{v\phi}_{jh}\right] \tag{A.5}$$

for $i = 1...4$ and $j, h = 1...8$, their non-null components read

$$\begin{aligned}
&P^{uu}_{11} = P^{uu}_{22} = P^{uu}_{33} = P^{uu}_{44} = 2\\
&P^{uu}_{55} = P^{uu}_{66} = P^{uu}_{77} = P^{uu}_{88} = 96\eta^3\varrho^2\\
&P^{vv}_{11} = P^{vv}_{22} = P^{vv}_{33} = P^{vv}_{44} = 96\varrho^2
\end{aligned} \tag{A.6}$$

$$\begin{aligned}
&P^{vv}_{55} = P^{vv}_{66} = P^{vv}_{77} = P^{vv}_{88} = 2\eta\\
&P^{\phi\phi}_{11} = P^{\phi\phi}_{22} = P^{\phi\phi}_{33} = P^{\phi\phi}_{44} = 8\varrho^2\\
&P^{\phi\phi}_{55} = P^{\phi\phi}_{66} = P^{\phi\phi}_{77} = P^{\phi\phi}_{88} = 8\eta\varrho^2\\
&P^{u\phi}_{55} = P^{u\phi}_{66} = -P^{u\phi}_{77} = -P^{u\phi}_{88} = -24\eta^2\varrho^2\\
&P^{v\phi}_{11} = -P^{v\phi}_{22} = P^{v\phi}_{33} = -P^{v\phi}_{44} = 24\varrho^2\\
&C^{uu}_{11} = C^{uu}_{22} = C^{uu}_{33} = C^{uu}_{44} = -2\\
&C^{uu}_{15} = C^{uu}_{26} = C^{uu}_{37} = C^{uu}_{48} = -96\eta^3\varrho^2\\
&C^{vv}_{11} = C^{vv}_{22} = C^{vv}_{33} = C^{vv}_{44} = -96\varrho^2\\
&C^{vv}_{15} = C^{vv}_{26} = C^{vv}_{37} = C^{vv}_{48} = -2\eta\\
&C^{\phi\phi}_{11} = C^{\phi\phi}_{22} = C^{\phi\phi}_{33} = C^{\phi\phi}_{44} = 4\varrho^2\\
&C^{\phi\phi}_{15} = C^{\phi\phi}_{26} = C^{\phi\phi}_{37} = C^{\phi\phi}_{48} = 4\eta\varrho^2\\
&C^{u\phi}_{15} = C^{u\phi}_{26} = -C^{u\phi}_{37} = -C^{u\phi}_{48} = 24\eta^2\varrho^2\\
&C^{v\phi}_{11} = -C^{v\phi}_{22} = C^{v\phi}_{33} = -C^{v\phi}_{44} = -24\varrho^2\\
&C^{\phi u}_{11} = C^{\phi u}_{22} = -C^{\phi u}_{33} = -C^{\phi u}_{44} = -\delta\\
&C^{\phi u}_{15} = C^{\phi u}_{26} = -C^{\phi u}_{37} = -C^{\phi u}_{48} = -24\eta^2\varrho^2\\
&C^{\phi v}_{11} = -C^{\phi v}_{22} = C^{\phi v}_{33} = -C^{\phi v}_{44} = 24\varrho^2\\
&C^{\phi v}_{15} = -C^{\phi v}_{26} = C^{\phi v}_{37} = -C^{\phi v}_{48} = -\eta\delta
\end{aligned} \tag{A.7}$$

## AppendixB. Asymptotic and perturbation methods

### *AppendixB.1. Generalized Scott equation*

A variant of the Scott version of the Faà di Bruno's formula [43], for the $n$-th derivative of the composite function $g(f(t))$ is here generalized for a sufficiently differentiable function $g(\mathbf{f}(t))$, depending on the vector function $\mathbf{f}(t) = (f_1(t), ..., f_i(t), .., f_\ell(t))$. The $n$-th derivative with respect to the $t$-variables reads

$$\frac{d^n}{dt^n} g\big(\mathbf{f}(t)\big) = \sum_{k=0}^{n} \sum_{|p|=k} \frac{1}{k!} \frac{\partial^{|p|}}{\partial \mathbf{f}^p} g(\mathbf{f}) \left[\frac{d^n}{dt^n}\big(\mathbf{f}(t)\big)^p\right]_{\mathbf{f}(t)\Rightarrow\mathbf{0}} \tag{B.1}$$

where $p$ denotes a multi-index (with $|p| = k$, and $k = 0, ..., n$) and $\mathbf{f}(t) \Rightarrow \mathbf{0}$ requires zeroing the function $\mathbf{f}(t)$ after formal $t$-differentiation. Denoting $G^{(n)}_0$ the $n$-th derivative evaluated in the generic value $t_0$ of the variable $t$, the following relation holds

$$G^{(n)}_0 = \left[\frac{d^n}{dt^n} g\big(\mathbf{f}(t)\big)\right]_{t=t_0} = \sum_{k=0}^{n} \sum_{|p|=k} \frac{1}{k!} \left[\frac{\partial^{|p|}}{\partial \mathbf{f}^p} g(\mathbf{f})\right]_{\mathbf{f}=\mathbf{f}_0} \left[\frac{d^n}{dt^n}\big(\mathbf{f}(t)\big)^p\right]_{\substack{\mathbf{f}(t)\Rightarrow\mathbf{0}\\ t=t_0}} \tag{B.2}$$

where $\mathbf{f}_0$ stands for the value assumed by $\mathbf{f}(t)$ in $t = t_0$ and the power $(\mathbf{f}(t))^p = f_{p_1}(t)...f_{p_k}(t)$, where $|p| = k$ with $k \in \mathbb{Z}^+$.

If the vector function $\mathbf{f}(t)$ is an integer $t$-power series function which can be expressed as $\mathbf{f}(t) = \sum \mathbf{f}^{[j]}\,(t-t_0)^j$, with $i$-th entry $f_i(t) = \sum f^{[j]}_i\,(t-t_0)^j$ and $i = 1..\ell$, then $\mathbf{z}_p(t) = (\mathbf{f}(t))^p$ is again an integer $t$-power series (of higher order) which can be expressed as $\mathbf{z}_p(t) = \sum \mathbf{z}^{[j]}_p\,(t-t_0)^j$. Therefore the notable relations yield

$$\left[\frac{d^n}{dt^n}\big(\mathbf{f}(t)\big)^p\right]_{\substack{\mathbf{f}(t)\Rightarrow\mathbf{0}\\ t=t_0}} = n!\left[\mathbf{z}^{[n]}_p\right]_{\mathbf{f}^{[0]}\Rightarrow\mathbf{0}}, \qquad \mathbf{f}_0 = \mathbf{f}^{[0]} \tag{B.3}$$

stating that the $n$-derivative of the $\mathbf{z}_p(t)$-series (if evaluated at $t=t_0$) depends only on the coefficient $\mathbf{z}_p^{[n]}$ of the $(t-t_0)^n$-power. This coefficient can be assessed by the recursive formula

$$\mathbf{z}_p^{[n]} = \sum_{i=1}^{\ell} \frac{1}{n f_i^{[0]}} \sum_{j=1}^{n} \Big( j(1+|p|)-n \Big) f_i^{[j]}\, \mathbf{z}_p^{[n-j]} \tag{B.4}$$

to be inizialized with $\mathbf{z}_p^{[0]} = (\mathbf{f}^{[0]})^p$. Finally, the $n$-th derivative for the composite function $g(\mathbf{f}(t))$ of a multi-variable integer-power series $\mathbf{f}(t)$, evaluated in $t_0$, reads

$$G_0^{(n)} = \sum_{k=0}^{n} \sum_{|p|=k} \frac{n!}{k!} \left[\mathbf{z}_p^{[n]}\right]_{\mathbf{f}^{[0]}\Rightarrow \mathbf{0}} \left[\frac{\partial^{|p|}}{\partial \mathbf{f}^p} g(\mathbf{f})\right]_{\mathbf{f}=\mathbf{f}^{[0]}} \tag{B.5}$$

and can be applied to directly obtain the $n$-th derivative $G_0^{(n)}$ in the *MPPM* expansion (39) of the characteristic function (where the zero-subscript is omitted for the sake of simplicity) by assuming $\epsilon$ as independent $t$-variable, $\boldsymbol{\nu}(\epsilon)$ as interior function $\mathbf{f}(t)$ and $F(\boldsymbol{\nu}(\epsilon))$ as exterior function $g(\mathbf{f}(t))$. Similarly, the $n$-th derivative $G_0^{(n)}$ in the *SPPM* expansion (28) of the characteristic function can be obtained as special sub-case, by assuming $\xi$ as independent $t$-variable, and defining the two-component vector variable $\mathbf{f}(t) = (\lambda(\xi), \xi)$.

The general formula can be specialized to deal with the $n$-th derivative of the two-variable composite function $g(\alpha(t), \mathbf{x}(t))$, depending on the distinct scalar $\alpha(t)$ and vector function $\mathbf{x}(t) = (x_1(t), ..., x_\ell(t))$. The $n$-th derivative reads

$$\frac{d^n}{dt^n} g\Big(\alpha(t), \mathbf{x}(t)\Big) = \tag{B.6}$$
$$= \sum_{\mathcal{S}(h,k)} \sum_{|p|=k} \frac{1}{(h+k)!} D^{h,|p|}_{g(\alpha,\mathbf{x})} \left[\frac{d^n}{dt^n}\Big(\big(\alpha(t)\big)^h \big(\mathbf{x}(t)\big)^p\Big)\right]_{\substack{\alpha(t)\Rightarrow 0\\ \mathbf{x}(t)\Rightarrow \mathbf{0}}}$$

where the index set $\mathcal{S}(h,k) = \Big\{h, k \in [0, h+k=n] \cup h, k \in \mathbb{Z}^+\Big\}$ and

$$D^{h,|p|}_{g(\alpha,\mathbf{x})} = \frac{\partial^{h+|p|}}{\partial \alpha^h \partial \mathbf{x}^p} g(\alpha, \mathbf{x}) = \frac{\partial^{h+|p|} g(\alpha, x_{p_1}, ..., x_{p_k})}{\partial \alpha^h \partial x_{p_1} ... \partial x_{p_k}} \tag{B.7}$$

If the $n$-th derivative is evaluated in the generic value $t_0$ of the independent variable $t$, the following relation holds

$$G_0^{(n)} = \frac{d^n}{dt^n}\Big[g\Big(\alpha(t), \mathbf{x}(t)\Big)\Big]_{t=0} = \tag{B.8}$$
$$= \sum_{\mathcal{S}(h,k)} \sum_{|p|=k} \frac{1}{(h+k)!} \left[D^{h,|p|}_{g(\alpha,\mathbf{x})}\right]_{\substack{\alpha=\alpha_0\\ \mathbf{x}=\mathbf{x}_0}} \left[\frac{d^n}{dt^n}\Big(\big(\alpha(t)\big)^h \big(\mathbf{x}(t)\big)^p\Big)\right]_{\substack{\alpha(t)\Rightarrow 0\\ \mathbf{x}(t)\Rightarrow \mathbf{0}\\ t=t_0}}$$

where $\alpha_0$ and $\mathbf{x}_0$ stands for the values of $\alpha(t)$ and $\mathbf{x}(t)$ in $t=t_0$.

If both the interior functions can be expressed as $t$-power series $\alpha(t) = \sum \alpha^{[j]}(t-t_0)^j$ and $\mathbf{x}(t) = \sum \mathbf{x}^{[j]}(t-t_0)^j$, with $i$-th entry $x_i(t) = \sum x_i^{[j]}(t-t_0)^j$, then $\mathbf{y}_{hp}(t) = (\alpha(t))^h(\mathbf{x}(t))^p$ is again a $t$-power series (of higher order) which can be expressed as $\mathbf{y}_{hp}(t) = \sum \mathbf{y}_{hp}^{[j]}(t-t_0)^j$. Therefore $\alpha_0 = \alpha^{[0]}$, $\mathbf{x}_0 = \mathbf{x}^{[0]}$ and

$$\left[\frac{d^n}{dt^n}\Big(\big(\alpha(t)\big)^h \big(\mathbf{x}(t)\big)^p\Big)\right]_{\substack{\alpha(t)\Rightarrow 0\\ \mathbf{x}(t)\Rightarrow \mathbf{0}\\ t=t_0}} = n!\left[\mathbf{y}_{hp}^{[n]}\right]_{\substack{\alpha^{[0]}\Rightarrow 0\\ \mathbf{x}^{[0]}\Rightarrow \mathbf{0}}} \tag{B.9}$$

where the coefficient $\mathbf{y}_{hp}^{[n]}$ obeys to the recursive formula

$$\mathbf{y}_{hp}^{[n]} = \sum_{i=1}^{\ell} \frac{1}{n\,\alpha^{[0]} x_i^{[0]}} \sum_{j=1}^{n} \Big(j(1+h+|p|)-n\Big)\, \alpha^{[j]} x_i^{[j]}\, \mathbf{y}_{hp}^{[n-j]} \tag{B.10}$$

to be initialized with $\mathbf{y}_{hp}^{[0]} = \left(\alpha^{[0]}\right)^h \left(\mathbf{x}^{[0]}\right)^p$. Finally, the $n$-th derivative for the composite function $g(\mathbf{f}(t))$ of a multi-variable integer-power series $\mathbf{f}(t)$, evaluated in $t_0$, reads

$$G_0^{(n)} = \sum_{\mathcal{S}(h,k)} \sum_{|p|=k} \frac{n!}{(h+k)!} \left[D^{h,|p|}_{g(\alpha,\mathbf{x})}\right]_{\substack{\alpha=\alpha^{[0]}\\ \mathbf{x}=\mathbf{x}^{[0]}}} \left[\mathbf{y}_{hp}^{[n]}\right]_{\substack{\alpha^{[0]}\Rightarrow 0\\ \mathbf{x}^{[0]}\Rightarrow \mathbf{0}}} \tag{B.11}$$

and can be applied to directly obtain the $n$-th coefficient $G^{(n)}$ of the *MPPM* equation (43) (where the subscript is omitted) by assuming $\epsilon$ as independent $t$-variable, $\lambda(\epsilon)$ and $\boldsymbol{\mu}(\epsilon)$ as interior functions $\alpha(t)$ and $\mathbf{x}(t)$ and finally $F(\lambda(\epsilon), \boldsymbol{\mu}(\epsilon))$ as exterior function $g(\alpha(t), \mathbf{x}(t))$. Similarly, the $n$-th derivative $G_0^{(n)}$ in the *SPPM* expansion (28) of the characteristic function can be obtained as special sub-case, by assign $\xi$ the twofold role of independent $t$-variable and single-component vector variable $\mathbf{x}(t)$, and then defining $\lambda(\xi)$ as interior scalar function $\alpha(t)$.

*AppendixB.2. SPPM asymptotic approximation*

The explicit parametric expressions of the twelve eigenvalues of the ideal system (for $\beta_1^* = 0$) in Table 4 must be distinguished depending on the aspect ratio. First, for the square cell ($\eta = 1$) the twelve zeroth-order eigenvalues are

$$\lambda_1^* = \lambda_2^* = 0, \quad \lambda_3^* = \lambda_4^* = 4, \tag{B.12}$$
$$\lambda_5^* = \lambda_6^* = 48\varrho^2, \quad \lambda_7^* = \lambda_8^* = 4 + 48\varrho^2$$
$$\lambda_9^* = \lambda_{10}^* = \frac{16\varrho^2 + \delta^2}{\chi^2}, \quad \lambda_{11}^* = 24\frac{\varrho^2}{\chi^2}, \quad \lambda_{12}^* = 2\frac{4\varrho^2 + \delta^2}{\chi^2}$$

whereas the coefficients of the second-order terms are

$$\mathcal{Y}_{12} = \frac{3}{2}\varrho^2 \tag{B.13}$$
$$\mathcal{Y}_{22} = \frac{1}{8}\frac{\delta^2 + 8\varrho^2}{\delta^2 + 4\varrho^2}$$
$$\mathcal{Y}_{32} = \frac{1}{4}\frac{2\left(12\chi^2 - 3\delta^2 + 4\right)\varrho^2 - 24\varrho^4 - 2\chi^2 + \delta^2 + \mathcal{R}_{342}}{4\chi^2 - \delta^2 - 16\varrho^2}$$
$$\mathcal{Y}_{42} = \frac{1}{4}\frac{2\left(12\chi^2 - 3\delta^2 + 4\right)\varrho^2 - 24\varrho^4 - 2\chi^2 + \delta^2 - \mathcal{R}_{342}}{4\chi^2 - \delta^2 - 16\varrho^2}$$
$$\mathcal{Y}_{52} = \frac{1}{2}\frac{\left(12\left(7 - 12\chi^2\right)\varrho^2 + 12\chi^2 + 3\delta^2 - 4 + \mathcal{R}_{452}\right)\varrho^2}{\left(48\chi^2 - 16\right)\varrho^2 - \delta^2}$$
$$\mathcal{Y}_{62} = \frac{1}{2}\frac{\left(12\left(7 - 12\chi^2\right)\varrho^2 + 12\chi^2 + 3\delta^2 - 4 - \mathcal{R}_{452}\right)\varrho^2}{\left(48\chi^2 - 16\right)\varrho^2 - \delta^2}$$
$$\mathcal{Y}_{72} = -\frac{1}{16}\frac{48\chi^2\varrho^2 + 4\chi^2 - \delta^2 - 24\varrho^2}{12\chi^2\varrho^2 + \chi^2 - 6\varrho^2}$$
$$\mathcal{Y}_{82} = -3\frac{\left(24\chi^2\varrho^2 + 2\chi^2 - \delta^2 - 10\varrho^2\right)\varrho^2}{24\chi^2\varrho^2 + 2\chi^2 - \delta^2 - 4\varrho^2}$$
$$\mathcal{Y}_{92} = \frac{64\left(3\chi^2 + 2\right)\varrho^4 + 8\left(6\chi^2 - 1\right)\delta^2\varrho^2 + 4\chi^2\delta^2 - \delta^4}{16\left(\delta^2 + 16\varrho^2 - 48\chi^2\varrho^2\right)\chi^2}$$

$$\mathcal{Y}_{102} = \frac{64\left(9\chi^2-2\right)\varrho^4+8\left(4\chi^2+\delta^2\right)\varrho^2+\delta^4}{16\,(\delta^2+16\varrho^2-4\chi^2)\,\chi^2}$$

$$\mathcal{Y}_{112} = \frac{\left(24\left(6\chi^4-5\chi^2+1\right)\varrho^2+2\chi^2(6\chi^2+3\delta^2-2)-3\delta^2\right)\varrho^2}{8\,(6\,(2\chi^2-1)\,\varrho^2+\chi^2)\,\chi^2}$$

$$\mathcal{Y}_{122} = \frac{8\left(3\chi^2+1\right)\varrho^6+\left(12\chi^2\delta^2-4\chi^2+3\delta^2\right)\varrho^4}{(\delta^2+4\varrho^2)\,(\delta^2+4\varrho^2-24\chi^2\varrho^2-2\chi^2)\,\chi^2}+ \\ +\frac{24\left(2\chi^4\delta^2-\chi^2\delta^4\right)\varrho^2+4\chi^4\delta^2-4\chi^2\delta^4+\delta^6}{16\,(\delta^2+4\varrho^2)\,(24\chi^2\varrho^2+2\chi^2-\delta^2-4\varrho^2)\,\chi^2}$$

where the auxiliary $\mathcal{R}$-coefficients read

$$\mathcal{R}_{342} = \Big(576\varrho^8+96\left(3\delta^2-12\chi^2+4\right)\varrho^6+ \\ 4(144\chi^4-72\chi^2\delta^2+9\delta^4-120\chi^2+72\delta^2+16)\varrho^2+ \\ 4(24\chi^2-18\chi^2\delta^2+3\delta^2-8\chi^2+4\delta^2)\varrho^2+(\delta^2-\chi^2)^2\Big)^{1/2} \tag{B.14}$$

$$\mathcal{R}_{452} = \Big(144(144\chi^4-168\chi^2+49)\varrho^4+ \\ 24(144\chi^4-36\chi^2\delta^2-132\chi^2+21\delta+28)\varrho^2+ \\ 9\delta^4+4(15-18\chi^2)+144\chi^4-96\chi^2+16\Big)^{1/2}$$

Second, for the rectangular cell ($\eta \neq 1$) the twelve zeroth-order eigenvalues are

$$\lambda_1^* = \lambda_2^* = 0, \quad \lambda_3^* = 4, \quad \lambda_4^* = 48\varrho^2, \quad \lambda_5^* = 4\eta, \tag{B.15}$$

$$\lambda_6^* = 48\eta^3\varrho^2, \quad \lambda_7^* = 4+48\eta^3\varrho^2, \quad \lambda_8^* = 4\eta+48\varrho^2$$

$$\lambda_9^* = 12\frac{\varrho^2(1+\eta)}{\chi^2}, \quad \lambda_{10}^* = \frac{\delta^2\eta+4\eta\varrho^2+\delta^2+4\varrho^2}{\chi^2},$$

$$\lambda_{11}^* = \frac{12\eta\varrho^2+\delta^2+4\varrho^2}{\chi^2}, \quad \lambda_{12}^* = \frac{12\varrho^2+\delta^2\eta+4\eta\varrho^2}{\chi^2}$$

and the related coefficients of the second-order terms are

$$\mathcal{Y}_{12} = 3\frac{\eta\varrho^2}{\eta+1}$$

$$\mathcal{Y}_{22} = \frac{1}{4}\frac{\delta^2\eta+4\eta\varrho^2+4\varrho^2}{\delta^2\eta+4\eta\varrho^2+\delta^2+4\varrho^2}$$

$$\mathcal{Y}_{32} = -\frac{1}{4}\frac{\eta(\delta^2+4\varrho^2)-4\chi^2+\delta^2+12\varrho^2}{\eta(\delta^2+4\varrho^2)-4\chi^2+12\varrho^2}$$

$$\mathcal{Y}_{42} = -\frac{3\left(12(4\chi^2-\eta)\varrho^2-\delta^2-16\varrho^2\right)\varrho^2}{12(4\chi^2-\eta)\varrho^2-\delta^2-4\varrho^2}$$

$$\mathcal{Y}_{52} = \frac{3\left(4\chi^2-\delta^2-4\varrho^2\right)\eta\varrho^2}{(4\chi^2-\delta^2-4\varrho^2)\eta-12\varrho^2}$$

$$\mathcal{Y}_{62} = \frac{\left(12\chi^2\eta^3-3\eta-1\right)\varrho^2}{12(4\chi^2\eta^2-1)\eta\varrho^2-\delta^2-4\varrho^2}$$

$$\mathcal{Y}_{72} = -\frac{1}{16}\frac{12(4\chi^2\eta^3-\eta-1)\varrho^2+4\chi^2-\delta^2}{3(4\chi^2\eta^3-\eta-1)\varrho^2+\chi^2}$$

$$\mathcal{Y}_{82} = -\frac{3\left((4\chi^2-\delta^2-4\varrho^2)\eta+16(3\chi^2-1)\varrho^2-\delta^2\right)\varrho^2}{(4\chi^2-\delta^2-4\varrho^2)\eta+4(12\chi^2-1)\varrho^2-\delta^2}$$

$$\mathcal{Y}_{92} = \frac{\mathcal{N}_{94}\eta^4+\mathcal{N}_{93}\eta^3+\mathcal{N}_{92}\eta^2+\mathcal{N}_{91}\eta+\mathcal{N}_{90}}{16\,(\eta+1)\,(3(4\chi^2\eta^3-\eta-1)\varrho^2+\chi^2)\,\chi^2}$$

$$\mathcal{Y}_{102} = \frac{\mathcal{N}_{102}\eta^2+\mathcal{N}_{101}\eta+\mathcal{N}_{100}}{16\chi^2\,(\delta^2+4\varrho^2)\,(\eta+1)\,\mathcal{Q}_{102}}$$

$$\mathcal{Y}_{112} = \frac{\mathcal{N}_{114}\eta^4+\mathcal{N}_{113}\eta^3+\mathcal{N}_{112}\eta^2+\mathcal{N}_{111}\eta+\mathcal{N}_{110}}{16\chi^2\mathcal{Q}_{112}}$$

$$\mathcal{Y}_{122} = \frac{\mathcal{N}_{122}\eta^2+\mathcal{N}_{121}\eta+\mathcal{N}_{120}}{16\chi^2\mathcal{Q}_{112}}$$

where the auxiliary $\mathcal{N}$- and $\mathcal{Q}$-coefficients read

$$\mathcal{N}_{94} = 12\left(\delta^2-8\varrho^2\right)\chi^2\varrho^2$$

$$\mathcal{N}_{93} = 12\left(48\chi^2\varrho^2+\delta^2-8\varrho^2\right)\varrho^2\chi^2$$

$$\mathcal{N}_{92} = 24\varrho^4-3\delta^2\varrho^2$$

$$\mathcal{N}_{91} = 48\left(1-3\chi^2\right)\varrho^4-2\left(4\chi^2+3\delta^2\right)\varrho^2$$

$$\mathcal{N}_{90} = 24\left(1-6\chi^2\right)\varrho^4+\left(48\chi^4-8\chi^2-3\delta^2\right)\varrho^2$$

$$\mathcal{N}_{102} = -\left(\delta^2+4\varrho^2\right)\left(\delta^2-8\varrho^2\right)\left(4\chi^2-\delta^2-4\varrho^2\right)$$

$$\mathcal{N}_{101} = -256\left(3\chi^2+1\right)\varrho^6+\left(4\chi^2-(12\chi^2+3)\delta^2\right)\varrho^4- \\ -48\chi^2\delta^4\varrho^2+16\chi^4\delta^2-8\chi^2\delta^4+2\delta^6$$

$$\mathcal{N}_{100} = -128\left(6\chi^2+1\right)\varrho^6-48\left(8\chi^2+1\right)\delta^2\varrho^4- \\ -16\left(3\delta^4-(12\chi^2-1)\delta^2\right)\chi^2\varrho^2-4\chi^2\delta^4+\delta^6$$

$$\mathcal{N}_{114} = 576\left(\delta^2-8\varrho^2\right)\chi^2\varrho^4$$

$$\mathcal{N}_{113} = -48\left(\delta^2+4\varrho^2\right)\left(48\chi^2\varrho^2-\delta^2+8\varrho^2\right)\varrho^2\chi^2$$

$$\mathcal{N}_{112} = -144\left(\delta^2-8\varrho^2\right)\varrho^4$$

$$\mathcal{N}_{111} = 768\left(3\chi^2+1\right)\varrho^6+96\left(6\chi^2+1\right)\delta^2\varrho^4+24\left(2\chi^2-\delta^2\right)\delta^2\varrho^2$$

$$\mathcal{N}_{110} = 128\left(6\chi^2+1\right)\varrho^6+48\left(8\chi^2+1\right)\delta^2\varrho^4+ \\ +16\left(3\delta^2-12\chi^2+1\right)\chi^2\delta^2\varrho^2+4\chi^2\delta^4-\delta^6$$

$$\mathcal{N}_{122} = -\left(\delta^2+4\varrho^2\right)\left(\delta^2-8\varrho^2\right)\left(4\chi^2-\delta^2-4\varrho^2\right)$$

$$\mathcal{N}_{121} = 728\left(3\chi^2-1\right)\varrho^6+32\left(18\chi^2\delta^2-3\delta^2+16\chi^2\right)\varrho^4+ \\ +8\left(3\delta^4-2\chi^2\delta^2-16\chi^4\right)\varrho^2$$

$$\mathcal{N}_{120} = 1152\left(6\chi^2-1\right)\varrho^6+48\left(3\delta^2-48\chi^4+8\chi^2\right)\varrho^4$$

$$\mathcal{Q}_{102} = 48\chi^2\varrho^2+4\chi^2\eta-\delta^2\eta-4\eta\varrho^2-\delta^2-4\varrho^2$$

$$\mathcal{Q}_{112} = \left(4(12\chi^2-3\eta-1)\varrho^2-\delta^2\right)\left(4(12\chi^2\eta^3-3\eta-1)\varrho^2-\delta^2\right)$$

$$\mathcal{Q}_{122} = \left((4\chi^2-\delta^2-4\varrho^2)\eta-12\varrho^2\right)\left(4\chi^2-12\varrho^2-(\delta^2+4\varrho^2)\eta\right)$$

and the cumbersome forth-order terms are omitted for the sake of conciseness.

*AppendixB.3. MPPM asymptotic approximation*

The explicit parametric expressions of the twelve eigenvalues of the ideal system (for $\beta_1^\circ = 0$) in Table 5 are

$$\lambda_1^\circ = \lambda_2^\circ = \lambda_3^\circ = \lambda_4^\circ = 0, \quad \lambda_5^\circ = \lambda_6^\circ = 4, \quad \lambda_7^\circ = \lambda_8^\circ = 4\eta \tag{B.16}$$

$$\lambda_9^\circ = (\eta+1)\frac{4\varrho^2+\delta^2}{\chi^2}, \quad \lambda_{10}^\circ = 12(\eta+1)\frac{\varrho^2}{\chi^2}$$

$$\lambda_{11}^\circ = 4(\eta+3)\frac{\varrho^2}{\chi^2}+\eta\frac{\delta^2}{\chi^2}, \quad \lambda_{12}^\circ = \frac{16\varrho^2+\delta^2}{\chi^2}$$

whereas the coefficients (all evaluable as $O(1)$ according to the parameter ordering) in the higher order eigensensitivities read

$$
\begin{aligned}
\mathcal{J}_{22} &= \frac{\varrho^2}{\delta^2+4\varrho^2}+\frac{\eta}{4\mathcal{G}_1}\frac{\delta^2}{(\delta^2+4\varrho^2)}\\
\mathcal{J}_{32} &= \mathcal{J}_{82} = 48 \qquad \text{(B.17)}\\
\mathcal{J}_{42} &= 48\eta^3+\frac{\mathcal{G}_2\beta_1^2}{\delta^2+4\varrho^2\mathcal{G}_2}\\
\mathcal{J}_{52} &= -\frac{1}{4}-\frac{\delta^2}{4\eta(\delta^2+4\varrho^2)+16(3\varrho^2-\chi^2)}\\
\mathcal{J}^{\varrho}_{62} &= 48\eta^3,\\
\mathcal{J}^{\beta}_{62} &= -\frac{1}{4}+\frac{\delta^2}{16\chi^2-48\mathcal{G}_1\varrho^2}\\
\mathcal{J}_{92} &= \frac{\delta^2}{4\mathcal{G}_1(\delta^2+4\varrho^2)}-\frac{\delta^2-8\varrho^2}{16\chi^2}\\
\mathcal{J}_{102} &= \frac{3\mathcal{G}_1\varrho^2\delta^2}{48\mathcal{G}_1\varrho^2\chi^2-16\chi^4}-\frac{\varrho^2}{2\chi^2}\\
\mathcal{J}_{112} &= \frac{\delta^2}{16\mathcal{G}_4\varrho^2+4\eta\delta^2-16\chi^2}+\frac{\delta^2-8\varrho^2}{16\chi^2}\\
\mathcal{J}_{122} &= \frac{\delta^2}{16\mathcal{G}_2\varrho^2+4\delta^2}-\frac{\delta^2-8\varrho^2}{16\chi^2}\\
\mathcal{J}_{14} &= 3\frac{\eta}{\mathcal{G}_1}\\
\mathcal{J}_{24} &= -\frac{\chi^2\delta^2(\eta\delta^2+4\mathcal{G}_1\varrho^2)}{16\mathcal{G}_1^3(\delta^2+4\varrho^2)^3}-\frac{\eta\mathcal{G}_6\delta^4+8\mathcal{G}_7\delta^2\varrho^2+16\mathcal{G}_1^2\varrho^4}{192\mathcal{G}_1^2(\delta^2+4\varrho^2)^2}\\
\mathcal{J}_{34} &= 3\frac{\mathcal{G}_2\beta_1^2+48\mathcal{G}_3(\delta^2+4(\mathcal{G}_2+3)\varrho^2)}{\mathcal{G}_2\beta_1^2+48\mathcal{G}_3(\delta^2+4\mathcal{G}_2\varrho^2)}\\
\mathcal{J}^{\beta}_{44} &= \frac{\mathcal{G}_2(\mathcal{G}_2+3)\varrho^2\beta_1^2(\delta^2+4\mathcal{G}_2\varrho^2)^2-6\mathcal{G}_2^2\varrho^2\beta_1^2\delta^2\chi^2}{24(\delta^2+4\mathcal{G}_2\varrho^2)^3(\mathcal{G}_2\beta_1^2+48\mathcal{G}_3(\delta^2+4\mathcal{G}_2\varrho^2))}+\\
&\quad+\frac{48\mathcal{G}_2(\mathcal{G}_5-3\eta^3)\varrho^2\chi^2\delta^2-3\mathcal{G}_1\mathcal{G}_2^2\varrho^4\beta_1^2}{4(\delta^2+4\mathcal{G}_2\varrho^2)^2(\mathcal{G}_2\beta_1^2+48\mathcal{G}_3(\delta^2+4\mathcal{G}_2\varrho^2))}+\\
&\quad+\frac{(7\eta^3+6\mathcal{G}_3\eta+\mathcal{G}_5)\varrho^2}{\mathcal{G}_2\beta_1^2+48\mathcal{G}_3(\delta^2+4\mathcal{G}_2\varrho^2)}\\
\mathcal{J}^{\varrho}_{44} &= -\frac{36\eta\mathcal{G}_3(\mathcal{G}_2\beta_1^2\varrho^2+16\eta^2\delta^2\chi^2)+36\eta^3\mathcal{G}_2\varrho^2\beta_1^2}{(\delta^2+4\mathcal{G}_2\varrho^2)(\mathcal{G}_2\beta_1^2+48\mathcal{G}_3(\delta^2+4\mathcal{G}_2\varrho^2))}\\
\mathcal{J}^{\beta}_{54} &= \frac{\mathcal{G}_1\delta^2+4(\mathcal{G}_4\varrho^2-\chi^2)}{192(\eta\delta^2+4(\mathcal{G}_4\varrho^2-\chi^2))}+\frac{\mathcal{G}_1\delta^2(\delta^2+4\varrho^2)}{64(\eta\delta^2+4(\mathcal{G}_4\varrho^2-\chi^2))^2}\\
\mathcal{J}^{\delta}_{54} &= \frac{\chi^2\delta^2\beta_1^2}{16(\eta\delta^2+4(\mathcal{G}_4\varrho^2-\chi^2))^3}\\
\mathcal{J}^{\delta}_{64} &= \frac{\mathcal{G}_2\varrho^2\beta_1^2-576\eta^3\mathcal{G}_1\varrho^4}{256(3\mathcal{G}_1\varrho^2-\chi^2)^2}+\frac{\beta_1^2+576\eta^3\varrho^2}{768(3\mathcal{G}_1\varrho^2-\chi^2)}\\
\mathcal{J}^{\beta}_{64} &= \frac{9\mathcal{G}_1\varrho^2\delta^4+16(3\mathcal{G}_1\varrho^2-\chi^2)^3}{3072(3\mathcal{G}_1\varrho^2-\chi^2)^3}\\
\mathcal{J}_{74} &= \frac{3\eta(\delta^2+4(\varrho^2-\chi^2))}{12\varrho^2+\eta(\delta^2+4(\varrho^2-\chi^2))}\\
\mathcal{J}_{84} &= -3\frac{16\varrho^2+\mathcal{G}_1\delta^2+4\eta(\varrho^2-\chi^2)}{4\varrho^2+\mathcal{G}_1\delta^2+4\eta(\varrho^2-\chi^2)}\\
\mathcal{J}^{\varrho}_{94} &= \frac{36\varrho^2}{\mathcal{G}_1(\delta^2+4\varrho^2)-4\eta\chi^2}\\
\mathcal{J}^{\delta}_{94} &= \frac{\mathcal{G}_1\beta_1^2(\delta^2+4\varrho^2)(\delta^2-8\varrho^2+4\chi^2)-4\chi^2\delta^2\beta_1^2}{64\mathcal{G}_1^3(\delta^2+4\varrho^2)^3}\\
\mathcal{J}^{\beta}_{94} &= \frac{\mathcal{G}_1(\delta^2+4\varrho^2)(\delta^2-8\varrho^2)-16\chi^2\delta^2}{768\mathcal{G}_1\chi^2(\delta^2+4\varrho^2)}\\
\mathcal{J}^{\varrho}_{104} &= -\frac{4\mathcal{G}_1\varrho^2(\beta_1^2-72\eta^3\delta^2)+\eta\beta_1^2\delta^2}{128(3\mathcal{G}_1\varrho^2-\chi^2)^2}-\\
&\quad-\frac{\mathcal{G}_1(\beta_1^2+72\eta^3\delta^2-864\varrho^2)+288\chi^2}{96\mathcal{G}_1(3\mathcal{G}_1\varrho^2-\chi^2)}\\
\mathcal{J}^{\delta}_{104} &= -\frac{3\mathcal{G}_1\beta_1^2\delta^2\varrho^2}{1024(3\mathcal{G}_1\varrho^2-\chi^2)^3}\\
\mathcal{J}^{\beta}_{104} &= -\frac{3\mathcal{G}_1^2\varrho^4(\delta^2-8\varrho^2)}{256\chi^2(3\mathcal{G}_1\varrho^2-\chi^2)^2}\\
\mathcal{J}^{\varrho}_{114} &= \frac{36\varrho^2}{4\mathcal{G}_4\varrho^2+\eta(\delta^2-4\chi^2)}\\
\mathcal{J}^{\delta}_{114} &= -\frac{\beta_1^2(4\mathcal{G}_1\varrho^2+\eta\delta^2)}{64(4\mathcal{G}_4\varrho^2+\eta\delta^2-4\chi^2)^2}-\frac{\beta_1^2\delta^2(4\mathcal{G}_1\varrho^2+\eta\delta^2)}{64(4\mathcal{G}_4\varrho^2+\eta\delta^2-4\chi^2)^3}\\
\mathcal{J}^{\beta}_{114} &= \frac{(4\mathcal{G}_1\varrho^2+\eta\delta^2)(\delta^2-8\varrho^2)+32\varrho^2\chi^2}{768\chi^2(4\mathcal{G}_4\varrho^2+\eta\delta^2-4\chi^2)}\\
\mathcal{J}^{\varrho}_{124} &= \frac{36\varrho^2}{4\mathcal{G}_2\varrho^2+\delta^2}\\
\mathcal{J}^{\delta}_{124} &= \frac{12\eta^3\varrho^2\chi^2}{(4\mathcal{G}_2\varrho^2+\delta^2)^2}-\frac{\beta_1^2\delta^2+8(2\mathcal{G}_2+3)\varrho^2\beta_1^2}{192(4\mathcal{G}_2\varrho^2+\delta^2)^2}\\
\mathcal{J}^{\beta}_{124} &= \frac{\mathcal{G}_2\varrho^2\delta^2\chi^2}{4(4\mathcal{G}_2\varrho^2+\delta^2)^3}+\frac{\delta^2-8\varrho^2}{768\chi^2} \qquad \text{(B.18)}
\end{aligned}
$$

where the auxiliary $\mathcal{G}$-parameters

$$
\begin{aligned}
&\mathcal{G}_1=\eta+1, \qquad \mathcal{G}_2=3\eta+1, \qquad \mathcal{G}_3=\eta^3-1, \qquad \text{(B.19)}\\
&\mathcal{G}_4=\eta+3, \qquad \mathcal{G}_5=\eta^3+1
\end{aligned}
$$

depend only on the cell geometric shape.